\documentclass{article}
\usepackage{version}
\includeversion{EXT}
\excludeversion{CODASPY}

\usepackage[table]{xcolor}
\colorlet{lightgray}{gray!20}
\colorlet{tablegray}{gray!60}

\usepackage{geometry}
\usepackage{mathtools}
\usepackage{enumitem}
\usepackage{booktabs}
\usepackage{array}
\usepackage[]{authblk}
\usepackage{empheq}
\usepackage{amssymb}
\usepackage{multicol}
\usepackage{subfig}
\usepackage{multirow}
\usepackage{wasysym}
\usepackage{algorithm}
\usepackage{algpseudocode}
\usepackage{tikz}
\usetikzlibrary{arrows,shapes,automata,positioning,petri,calc,scopes,calc}

\tikzset{auto, font=\scriptsize, thin, >= stealth}

\tikzstyle{line} = [draw, rounded corners, -latex']
\tikzstyle{Task} = [draw, minimum height=20pt,minimum width=20pt, fill=lightgray,align=center]
\tikzstyle{Block} = [draw, minimum height=20pt,minimum width=20pt, fill=lightgray,align=center]
\tikzstyle{RelativeLink} = [line, densely dashdotted, -triangle 45]
\tikzstyle{BranchingPoint} = [diamond, draw, minimum height=20pt,minimum width=20pt, fill=lightgray,align=center]
\tikzstyle{Link} = [line, rounded corners, ->]
\tikzstyle{ContingentLink} = [line, double, ->]

\newtheorem{definition}{Definition}
\newtheorem{theorem}{Theorem}
\newtheorem{lemma}{Lemma}
\newtheorem{proof}{Proof}

\newtheorem{securitypolicy}{Security Policy}

\newcommand{\A}{\mathcal{A}}
\newcommand{\T}{\mathcal{T}}
\newcommand{\C}{\mathcal{C}}
\newcommand{\Y}{\mathcal{Y}}

\newcommand{\Users}{\texttt{Users}}
\newcommand{\Roles}{\texttt{Roles}}
\newcommand{\Perm}{\texttt{Perm}}
\newcommand{\Sessions}{\texttt{Sess}}
\newcommand{\UA}{\mathit{UA}}
\newcommand{\PA}{\mathit{PA}}
\newcommand{\user}{\mathit{user}}
\newcommand{\role}{\mathit{role}}
\newcommand{\RH}{\mathit{RH}}

\newcommand{\ST}{\mathit{ST}}
\newcommand{\Enabled}{\texttt{enabled}}
\newcommand{\Disabled}{\texttt{disabled}}
\newcommand{\Active}{\texttt{active}}
\newcommand{\Inactive}{\texttt{inactive}}
\newcommand{\Enable}{\texttt{enable}}
\newcommand{\Disable}{\texttt{disable}}

\newcommand{\Cal}{\texttt{C}}
\newcommand{\Hours}{\mathit{Hours}}
\newcommand{\Days}{\mathit{Days}}
\newcommand{\Weeks}{\mathit{Weeks}}

\newcommand{\ptToSTN}{\mu_\mathit{pt2stn}}
\newcommand{\con}{\mu_\mathit{con}}

\newcommand{\OutwardJourney}{\texttt{OutwardJourney}}
\newcommand{\ReturnJourney}{\texttt{ReturnJourney}}
\newcommand{\SystemCheck}{\texttt{SystemCheck}}
\newcommand{\SecurityCheck}{\texttt{SecurityCheck}}
\newcommand{\TrainDriver}{\mathit{TrainDriver}}
\newcommand{\SecurityEngineer}{\mathit{SecurityEngineer}}
\newcommand{\SystemEngineer}{\mathit{SystemEngineer}}

\newcommand{\Alice}{\mathit{Alice}}
\newcommand{\Bob}{\mathit{Bob}}
\newcommand{\Charlie}{\mathit{Charlie}}
\newcommand{\Eve}{\mathit{Eve}}
\newcommand{\Kate}{\mathit{Kate}}
\newcommand{\wf}{\mathit{wf}}

\newsavebox{\mysaveboxM} 
\newsavebox{\mysaveboxT} 

\newcommand{\boxwidth}[1]{#1}
\newcommand{\SetBoxWidth}[1]{\renewcommand{\boxwidth}{#1}}

\newcommand*\Garybox[2][]{%
    \sbox{\mysaveboxM}{#2}%
    \sbox{\mysaveboxT}{\fcolorbox{black}{white}{#1}}%
    \sbox{\mysaveboxM}{%
      \parbox[b][\ht\mysaveboxM+.5\ht\mysaveboxT+.5\dp\mysaveboxT][b]{%
        \wd\mysaveboxM}{#2}%
    }%
    \sbox{\mysaveboxM}{%
      \fcolorbox{black}{white}{%
        \makebox[\boxwidth]{\usebox{\mysaveboxM}}%
      }%
}%
\usebox{\mysaveboxM}%
    \makebox[0pt][r]{%
      \makebox[\wd\mysaveboxM][c]{%
        \raisebox{\ht\mysaveboxM-0.5\ht\mysaveboxT
+0.5\dp\mysaveboxT-0.5\fboxrule}{\usebox{\mysaveboxT}}%
}%
}%
}

\begin{document}





\begin{CODASPY}
\title{Security Constraints in Temporal Role-Based Access-Controlled
Workflows\thanks{This work was partially supported by the PRIN 2010-2011
Project ``Security Horizons''. We thank the anonymous reviewers for their comments and suggestions.}}
\end{CODASPY}

\begin{EXT}
\title{Security Constraints in Temporal Role-Based Access-Controlled
Workflows \\ (Extended Version)\thanks{This work was partially supported by the PRIN 2010-2011
Project ``Security Horizons''. The short version of this paper will be published in the proceedings of CODASPY 2016. We thank the anonymous reviewers for their
comments and suggestions.}}
\end{EXT}

\author[1]{Carlo Combi\thanks{carlo.combi@univr.it}}
\author[2]{Luca Vigan\`o\thanks{luca.vigano@kcl.ac.uk}}
\author[1]{Matteo Zavatteri\thanks{matteo.zavatteri@univr.it}}
\affil[1]{Dipartimento di Informatica, Universit\`a di Verona, Italy}
\affil[2]{Department of Informatics, King's College London, UK}

\renewcommand\Authands{ and }

\date{}
\maketitle


\begin{abstract}

Workflows and role-based access control models need to be suitably merged, in
order to allow users to perform processes in a correct way, according to the
given data access policies and the temporal constraints. Given a mapping
between workflow models and simple temporal networks with uncertainty, we
discuss a mapping between role temporalities and simple temporal networks, and
how to connect the two resulting networks to make explicit \emph{who can do
what, when}. If the connected network is still executable, we show how to
compute the set of authorized users for each task. Finally, we define security
constraints (to prevent users from doing unauthorized actions) and security
constraint propagation rules (to propagate security constraints at runtime).
We also provide an algorithm to check whether a set of propagation rules is
safe, and we extend an existing execution algorithm to take into account these
new security aspects. 

%
\end{abstract}

%
%


\section{Introduction}




\textbf{Context and motivation.} Workflow technology has emerged as a key
technology to specify and manage business processes within complex
organizations. Recent research has focused on the issues related to workflow
temporalities, such as uncertain durations of tasks, temporal constraints
between (even non consecutive) tasks, deadlines and so on
\cite{DBLP:journals/tsmc/CombiGMP14}. As complex tasks need the suitable
access to data and systems, \emph{role-based access control (RBAC) models}
play an important part as they both deal with the classical security analysis
(concerning authorization inspection, administrative models, and hierarchies)
and allow one to consider also temporal
aspects~\cite{DBLP:conf/stm/ArmandoR10,DBLP:journals/tissec/BertinoBF01,DBLP:conf/IEEEias/MondalS08, DBLP:conf/sac/RaniseTV15,
DBLP:journals/computer/SandhuCFY96}.

Thus, in the business process context, RBAC models and workflow models need
to be suitably merged, in order to allow users to perform processes in a
correct way according to the given data access policies and the temporal constraints.

To properly manage temporal constraints of workflows, solutions have been
proposed that are based on a mapping between workflow models and \emph{simple
temporal constraint networks with uncertainty
(STNU)}~\cite{DBLP:journals/jetai/VidalF99} and that allow one to deal with
\emph{controllability} of workflow models. In a nutshell, an STNU, and its
corresponding workflow, is \emph{controllable} if it is always possible to
execute the network without violating any constraint no matter what the
uncertain durations (of tasks) turn out to
be~\cite{DBLP:journals/jetai/VidalF99}.

Even if both a temporal workflow model and a temporal access control model
pass their security analyses successfully, in general we cannot be sure that
their composition behaves as we expect. Hence, we need a way to analyze what
happens when we put an access control model on top of a workflow model. As a
temporal workflow can be translated into an equivalent STNU, if we were able
to extend this network to take into consideration the security aspects, we
could have some chances of reasoning on their interplay.

\textbf{Contributions.} In this direction, we have merged \emph{temporal
role-based access control (TRBAC)} and temporal workflow models to seamlessly
manage temporal constraints when executing tasks together with possible
temporal constraints related to the availability of agents able to execute
tasks according to their roles. 
Thus, the first two contributions of this paper are: 
proposing a mapping of valid intervals of roles into an equivalent simple temporal network, and 
merging it with the STNU specifying the workflow model.

The enabling times of the roles in TRBAC models are usually specified
according to periodic expressions using \emph{calendars}~\cite{niezette}. We
propose the concept of \emph{configuration}, which corresponds to an STNU
containing both the representation of a temporal workflow and the related
role-based access model, considering periodic role-enabling intervals within
a given, limited time window. A configuration allows us to check if the
workflow is executable with respect to the access control model. That is, it
allows us to understand if these two models are consistent with each other.
If so, then we are able to compute which users belonging to which roles are
authorized to execute the tasks.

Moreover, a further contribution is 
the definition of
\emph{security constraints (SCs)}, not directly expressible in role-based
access models, along with their \emph{security constraint propagation rules
(SCPRs)}. The former are used to prevent users from doing unauthorized
actions (e.g., starting/ending a task) if the current time satisfies the
constraint itself, whereas the latter are used to propagate these security
constraints depending on what is going on. If different users make different
choices, then SCPRs will propagate different SCs. This \emph{dynamic}
approach reacts to observations of the occurring runtime events. As far as we
know, this is the first attempt to use temporal networks to model (and
enforce) security (policies).

\textbf{Organization.} Section~\ref{sec:background}
reviews essential background of simple temporal networks (STNs), STNUs, a mapping from workflow models to STNUs, and TRBAC. Section~\ref{sec:acwf} provides a new mapping to
translate the enabling intervals of roles of TRBAC into an STN and a
connection mapping to connect the workflow STNU to the access control STN.
It also shows how to derive the set of authorized
users for each time point in these networks. Section~\ref{sec:caseStudy} introduces a case study also specifying
three security policies that are supposed to hold in that context.
Section~\ref{sec:scp} defines SCs and SCPRs to enforce (temporal) security policies when executing a temporal workflow. It also discusses a safeness algorithm for a set of SCPRs.
Section~\ref{sec:wfexec} discusses how to extend an already existing
execution algorithm for STNUs so as to take into account these rules too.
Section~\ref{sec:RW} discusses related work. Section~\ref{sec:CFW} draws conclusions and discusses future work. 
\begin{CODASPY}
The proofs of our results are given, along with further details, in~\cite{CombiViganoZavatteri15}.
\end{CODASPY}
\begin{EXT}
The proofs of our results are given in the appendix.
\end{EXT}

\section{Background}\label{sec:background}

In this section, we briefly review the theoretical foundations of
STNs~\cite{DBLP:conf/kr/DechterMP89} and STNUs~\cite{DBLP:conf/aaai/MorrisM05}, how to map a workflow into an STNU, and role based access models.

\subsection{Simple Temporal Networks}

\begin{definition}\em
A \emph{Simple Temporal Network (STN)} is a pair $\langle\T,\C\rangle$, where $\T$ is a set of time points with continuous domain, and $\C$ is a set of constraints of  the form $X - Y \leq k$ with $X,Y$ time points and $k \in \mathbb{R} \cup \{-\infty,\infty\}$~\cite{DBLP:conf/kr/DechterMP89}. 

A \emph{Simple Temporal Constraint Satisfaction problem (ST\-CP)} is the problem of finding a complete assignment of values to the time points in $\T$ 
satisfying all constraints in $\C$~\cite{DBLP:conf/kr/DechterMP89}.
\end{definition}

An STN can also be represented as a directed graph where each node
represents a time point of $\T$ and each edge $X \xrightarrow[]{[x,y]} Y$
(called \emph{requirement link} or \emph{link}, for short) represents the two
constraints $Y-X \leq y$ and $X-Y \leq -x$ belonging to $\C$. For each pair of
time points in a directed graph representing an STN, there exists only one
edge between them, which is labeled exactly by one range.
%
%
We can also represent it through an equivalent
directed weighted graph (called \emph{distance graph} $G_d$), where the set
of nodes is still the set of time points and each constraint $Y - X \leq k$ is
mapped to a weighted edge $X \xrightarrow[]{k} Y$. That is, each edge of the
STN $X \xrightarrow[]{[x,y]} Y$ is mapped to $X \xrightarrow[]{y} Y$ and $Y
\xrightarrow[]{-x} X$ in the distance graph.

To avoid confusion, hereinafter \emph{edges} will refer to the edges in the
distance graph, whereas \emph{(requirement) links} will refer to the edges in
the STN graph.


To find the ranges of distance values allowed between time points, one can
run the \emph{all pairs shortest paths algorithm} on the distance graph
$G_d$~\cite{Cormen:2009:IAT:1614191}. If $G_d$ contains a negative cycle,
the given STP does not admit solutions, i.e., it is \emph{inconsistent}.
The upper bound of the range between the $i^\mathrm{th}$ time point and the origin time point $Z$ corresponds to the shortest path from node $Z$ to that node, whereas the lower bound corresponds to the negation of the shortest path in the opposite direction.

Assuming that the origin time point $Z$ is the starting point, to find a complete solution $\texttt{S} = \{Z=0, X_1=t_{X_1}, \dots\}$ for each time point
$X$, we choose a value among those allowed in its range adding the link $Z
\xrightarrow[]{[x,x]} X$ to the STN (if a link $Z \to X$ already exists in
the STN graph, then we replace it with the new one). This translates into
adding $X - Z \leq x$ and $Z - X \leq -x$ to $\C$, which fixes the value for
$X$. To get the new updated ranges for the remaining time points, we
\emph{propagate} the effect of this assignment recomputing the shortest paths on the distance graph containing now the two new constraints $Z
\xrightarrow[]{x} X$ and $X \xrightarrow[]{-x} Z$. Managing in such way all time points, we obtain a complete solution. 
\subsection{STN with Uncertainty}

\begin{definition}\em
A \emph{Simple Temporal Network with Uncertainty (STNU)} extends an STN by adding a set of contingent links~\cite{DBLP:conf/ijcai/MorrisMV01}. Formally, an STNU is a triple $\mathcal{S} = \langle \T,\C,\mathcal{L} \rangle$ where: 
\begin{itemize}[noitemsep]
\item $\langle \T,\C \rangle$ is an STN,

\item $\mathcal{L}$ is a set of \emph{contingent links} of the form $(A,x,y,C)$ (or, equivalently, $A \xRightarrow[]{[x,y]} C$ in the STNU graph), where the \emph{activation} point $A$ and the \emph{contingent} point $C$ are different time points ($A\not\equiv C$), $x$ and $y$ are such that $0 < x < y <\infty$, and
\begin{itemize}[noitemsep]
\item for each $(A,x,y,C) \in \mathcal{L}$, $\C$ contains $C -A \leq y$ and $A -C \leq -x$,
\item if $(A_1,x_1,y_1,C_1)$ and $(A_2,x_2,y_2,C_2)$ are two distinct contingent links, then $C_1 \not\equiv C_2$,
\item the contingent time point of a contingent link may play the role of an activation point for another one.
\end{itemize}
\end{itemize}
When we are not interested in talking about the range of a link, we simply
write $A \Rightarrow C$ (for contingents) or $X \rightarrow Y$ (for
requirements) omitting $[x,y]$. 
As notation, we write $A$ to refer to activation time points, $C$ to
contingent time points and $X$ to generic time points. If $X$ is not a
contingent time point, then it is also called \emph{control time point}.
\end{definition}

It is easy to see that each STN $\langle \T, \C \rangle$ is also an STNU
$\langle \T, \C, \mathcal{L} \rangle$ where $\mathcal{L} = \emptyset$ (i.e.,
without contingent links). 

When the network is being executed, the system incrementally assigns a fixed
time value to each control time point (i.e., to each non-contingent time
point) among those allowed in its range. The system can only \emph{observe}
the occurrence of any contingent $C_i$, which is not under the control of the
system and is however guaranteed to occur in such a way that that $C_i -A_i
\in [x_i,y_i]$.

The meaning of contingency can be thought as representing processes
that are not under the control of the workflow systems and whose exact duration is unknown \emph{but} bounded by the range $[x,y]$.
For example, the writing of this paper once started (i.e., once $A$ has been
executed) will last at least a minimum amount of time $x$ to allow authors
to get a polished version to be submitted and at most $y$ $(> x)$, which in
this context is related to the submission deadline. However, the exact moment when the authors will have it finished (and consequently the paper will have
been submitted) is unknown at this stage.

In an STNU, we need to move from the concept of \emph{consistency} to that of
\emph{controllability}, because we now have to deal with ``uncertainty'', which
is by definition out of our control. We must make sure that no execution will
violate any constraint. Hence, an STNU is \emph{controllable} if we are able
to execute all control time points satisfying all constraints in $\C$ no
matter what the durations of the contingent links turn out to be. The rest of
this section provides everything we need to define the various types of
controllability (see~\cite{DBLP:journals/jetai/VidalF99} for details). 


A \emph{situation} $\omega = (d_1,\dots,d_n$) is defined by fixing a chosen
duration for each contingent link. Fixing a situation is equivalent to
transforming an STNU into an STN as each $A_i \xRightarrow[]{[x_i,y_i]} C_i$ is replaced with $A_i \xrightarrow[]{[d_i,d_i]} C_i$ ($d_i \in [x_i,y_i]$). A \emph{(situation) projection} is a
mapping $\mathit{sitPrj} : \langle \T,\C,\mathcal{L} \rangle \times \Omega \to
\langle \T,\C' \rangle$, which considers all contingent links as if they were
requirement links with a fixed distance. Therefore, an STNU represents an
infinite family of STNs (each one \emph{projecting} a different situation).
The \emph{space of all situations} is represented by $\Omega$. A
\emph{schedule} is a mapping $\psi : \T \to \mathbb{R}$ that assigns a real
value to each time point. The \emph{space of all schedules} (since an STNU
can be executed in infinite ways) is represented by $\Psi$. An
\emph{execution strategy} for $\mathcal{S}$ is a mapping $\sigma : \Omega \to
\Psi$ such that for each situation $\omega \in \Omega$, $\sigma(\omega)$ is a
complete schedule for the time points in $\T{}$.

Three main kinds of controllability for STNUs have been originally
defined in~\cite{ DBLP:journals/jetai/VidalF99}.
An STNU is \emph{weakly controllable} if there exists a viable execution
strategy, i.e., if every projection is consistent. An STNU is \emph{strongly controllable} if there exists a set of viable execution strategies considering all possible projections, where each control time point is assigned the same value by the schedule in all the strategies. An STNU is \emph{dynamically controllable} if there
exists an execution strategy for $\mathcal{S}$ that is both viable and
dynamic$^*$, where an execution strategy is dynamic$^*$ whenever: if the
durations of all contingent time points executed before the next time point
$X$ are equal in all different situations $\omega_1$ and $\omega_2$, then the schedule must assign the same value to $X$ in both $\omega_1$ and $\omega_2$.

Checking the dynamic controllability of an STNU is polynomial. The first
algorithm was proposed in~\cite{DBLP:conf/ijcai/MorrisMV01} and further
improvements were given in~\cite{DBLP:conf/cp/Morris06,DBLP:conf/aaai/MorrisM05}.
The main idea behind the controllability check is that
of restricting the execution strategies ruling out those that would squeeze
the contingent links, where a contingent link is \emph{squeezed} if the other constraints imply a tighter lower and/or upper bound for the link.
Hereinafter, we will refer to the basic algorithm introduced in~\cite{DBLP:conf/aaai/MorrisM05} for STNU dynamic controllability, avoiding the discussion related to subtle further optimizations.
The algorithm takes as input a labeled distance graph built from the STNU
according to the following mapping: each requirement link
$X \xrightarrow[]{[x,y]} Y$ is mapped to $X \xrightarrow[]{y} Y$ and $Y
\xrightarrow[]{-x} X$ in the labeled distance graph. For each contingent link
$A \xRightarrow[]{[x,y]} C$, we have the same edges $A \xrightarrow[]{y} C$ and $C \xrightarrow[]{-x} A$, but we also have $A \xrightarrow[]{c:x} C$ and $C \xrightarrow[]{C:-y} A$ which are the \emph{lower-case} and the \emph{upper-case} edge, respectively.





	


The algorithm proposed in~\cite{DBLP:conf/aaai/MorrisM05} iteratively checks if the
\emph{AllMax} projection (i.e., the projection in which all contingent links
take their maximal duration) is consistent in the STN-sense, where the
\emph{AllMax} projection is the unlabeled distance graph obtained by deleting
all lower-case edges and all labels from the upper-case edges (whenever we
remove labels from edges or we add new edges, if an edge of the same type
already exists in the graph we are operating on, then we usually keep that
specifying the tighter constraint with respect to the type of edge). If so, it
generates new edges according to suitable edge generation rules given in
\cite{DBLP:conf/aaai/MorrisM05}, until either quiescence (i.e., no further
constraints are added or the existing ones are tightened) or the cutoff bound
used to make the algorithm strongly polynomial, is reached. A detailed
analysis of the execution of this algorithm as well as how the proposed rules
work can be found in~\cite{DBLP:conf/aaai/MorrisM05}.

\subsection{Workflow Modeling}

A \emph{workflow} consists of a set of tasks to be executed in some order to
achieve some (business) goal(s). A \emph{temporal workflow} extends the
classical one by taking into account temporal constraints that typically
require a lower and an upper bound on the duration of tasks. A temporal
workflow also allows one to express relative constraints restricting the
allowed time distance between the start or the end of two (not necessarily
consecutive) tasks.

In this paper, we only consider \emph{structured} workflows that can be
described by a well-defined grammar and, without loss of generality, we do not
consider alternative/choice/con\-ditional paths. Thus, we will focus on the
workflow specification where all the specified tasks have to be properly
executed. An example of the basic constructs of this grammar is given in
Table~\ref{tbl:wf2stnu} (\textsc{Workflow Block}s), where, for each block,
the equivalent STNU is depicted on the right of it (\textsc{Equivalent STNU}).

The table shows the basic workflow block task (first row), which can be
thought of as a terminal symbol, and then the sequence (second row) and
parallel (third row), which can be thought of as non-terminal symbols. The
last component (relative constraint) in the fourth row only imposes further
temporal constraints between the start/end of two tasks and it has nothing to
do with the control flow that is regulated by the grammar. If the
workflow model is structured, we will ``structure'' the corresponding STNU.

\begin{table}[t]
\caption{Workflow to STNU mapping.}
\label{tbl:wf2stnu}
\centering 
\begin{tabular}{>{\centering\arraybackslash}m{6cm}>{\centering\arraybackslash}m{6cm}}
\toprule
\textsc{Workflow block} & \textsc{Corresponding STNU} \\
\midrule
\begin{tikzpicture}[auto,node distance=30pt]
			\node[Task,label={below:$[x,y]$}] (T) {$T$};
\end{tikzpicture}
& 
\begin{tikzpicture}[auto,node distance=30pt]
			\node (A) {$A$};
			\node (C) [right=of A] {$C$};
			\draw[ContingentLink]  (A) to node {} node [] {$[x,y]$} (C);
\end{tikzpicture}\\
\midrule
\begin{tikzpicture}[auto,node distance=30pt]
			\node[Block] (WFBLOCK1) {$\textsc{WF-Block}_1$};
			\node[Block] [right=of WFBLOCK1,xshift=-5pt] (WFBLOCK2) {$\textsc{WF-Block}_2$};
			\draw[Link]  (WFBLOCK1) to node {} node [] {$[x,y]$} (WFBLOCK2);
\end{tikzpicture}

& 
\begin{tikzpicture}[auto,node distance=30pt]
			\node (WFBLOCK1E) {$\textsc{WFB}_1^E$};
			\node (WFBLOCK2S) [right=of WFBLOCK1E] {$\textsc{WFB}_2^S$};
			\draw[Link]  (WFBLOCK1E) to node {} node [] {$[x,y]$} (WFBLOCK2S);			
\end{tikzpicture}\\

\multicolumn{2}{p{12cm}}{$\textsc{WFB}_2^S$ (resp., $\textsc{WFB}_1^E$) is a convention to represent the start (resp., the end) time point of the workflow block $\textsc{WF-Block}_2$ (resp., $\textsc{WF-Block}_1$).}\\

\midrule 
\begin{tikzpicture}[auto,node distance=30pt]
			\node[BranchingPoint,label={[xshift=-5pt]below right:$[x,y]$}] (B) {$=$};
			\node[BranchingPoint,label={[xshift=-5pt]below right:$[z,k]$}] (E) [right=of B,xshift=30pt] {$=$};
			
			\node[Block] (Bdot) [right=of B,xshift=-10pt] {$\dots$};
			\node[Block] (WFBLOCK1) [above=of Bdot,yshift=-20pt] {$\textsc{WF-Block}_1$};
			\node[Block] (WFBLOCKn) [below=of Bdot,yshift=20pt] {$\textsc{WF-Block}_n$};
			
			\draw[Link]  (B) |- node[] {$[a,b]$}  (WFBLOCK1);
			\draw[line, loosely dotted]  (B) -- (Bdot);
			\draw[Link]  (B) |-  node[swap] {$[c,d]$}  (WFBLOCKn);

			\draw[Link]  (WFBLOCK1) -|  node[] {$[e,f]$}  (E);
			\draw[line, loosely dotted]  (Bdot) -- (E);
			\draw[Link]  (WFBLOCKn) -|  node[swap] {$[g,h]$} (E);
\end{tikzpicture}

& 
\begin{tikzpicture}[auto,node distance=30pt]
			
			\node (BS) {$B^S$};
			\node (BE) [below=of BS,yshift=20pt] {$B^E$};

			\node (WFBLOCKdotS) [below=of BE,yshift=20pt] {$\textsc{WFB}_{\dots}^S$};
			\node (WFBLOCKdotE) [below=of WFBLOCKdotS,yshift=20pt] {$\textsc{WFB}_{\dots}^E$};

			\node (WFBLOCK1S) [left=of WFBLOCKdotS,xshift=20pt] {$\textsc{WFB}_1^S$};
			\node (WFBLOCK1E) [below=of WFBLOCK1S,yshift=20pt] {$\textsc{WFB}_1^E$};

			\node (WFBLOCKnS) [right=of WFBLOCKdotS,xshift=-20pt] {$\textsc{WFB}_n^S$};
			\node (WFBLOCKnE) [below=of WFBLOCKnS,yshift=20pt] {$\textsc{WFB}_n^E$};

			\node (ES) [below=of WFBLOCKdotE,yshift=20pt]{$E^S$};
			\node (EE) [below=of ES,yshift=20pt] {$E^E$};

			\draw[Link]  (BS) to node {} node [] {$[x,y]$} (BE);		
			\draw[Link]  (BE) to node {} node [swap] {$[a,b]$} (WFBLOCK1S);
			\draw[line, loosely dotted]  (BE) to node {} node [] {} (WFBLOCKdotS);
			\draw[Link]  (BE) to node {} node [] {$[c,d]$} (WFBLOCKnS);
			
			\draw[line, loosely dotted]  (WFBLOCK1S) to node {} node [] {} (WFBLOCK1E);
			\draw[line, loosely dotted]  (WFBLOCKdotS) to node {} node [] {} (WFBLOCKdotE);
			\draw[line, loosely dotted]  (WFBLOCKnS) to node {} node [] {} (WFBLOCKnE);
			
			\draw[Link]  (WFBLOCK1E) to node {} node [swap] {$[e,f]$} (ES);
			\draw[line, loosely dotted]  (WFBLOCKdotE) to node {} node [] {} (ES);
			\draw[Link]  (WFBLOCKnE) to node {} node [] {$[g,h]$} (ES);
			
			\draw[Link]  (ES) to node {} node [] {$[z,k]$} (EE);		
\end{tikzpicture}\\

\multicolumn{2}{p{12cm}}{$B^S \to B^E$ (resp., $E^S \to E^E$) is a convention to represent the branch (resp., join) component as an internal task.}\\

\midrule 
\begin{tikzpicture}[auto,node distance=30pt]
			\node[Block] (Bi) {$\textsc{WF-Block}_i$};
			\node[Block] (Bj)[right=of Bi,xshift=-5pt] {$\textsc{WF-Block}_j$};
			\draw[line, loosely dotted]  (Bi) to node {} node [] {} (Bj);
			\draw[RelativeLink] (Bi.north) -- ++  (0pt,10pt)  -| node[near start] {$\langle I_S \rangle[x,y]\langle I_F \rangle$} (Bj);
\end{tikzpicture}

& 
\begin{tikzpicture}[auto,node distance=30pt]
			\node (SEBi) {$X$};
			\node (SEBj) [right=of SEBi] {$Y$};
			\draw[line, loosely dotted]  (SEBi) to node {} node [] {} (SEBj);
			\draw[Link] (SEBi.north) -- ++  (0pt,10pt)  -| node[near start] {$[x,y]$} (SEBj);
\end{tikzpicture}\\

\multicolumn{2}{p{12cm}}{$X$ is either an activation point (if $\langle I_S
\rangle \equiv S$) or a contingent point (if $\langle I_S \rangle \equiv E$)
of some task inside $\textsc{WF-Block}_i$ and $Y$ is the same but with respect to $\textsc{WF-Block}_j$.} \\

\bottomrule
\end{tabular}
\end{table}

\subsection{Temporal RBAC Models}\label{sec:trbacmodels}

So far, we have not talked about workflow security, but, as we mentioned
above, our aim is to put a temporal \emph{Role-Based Access Control (RBAC, \cite{DBLP:journals/computer/SandhuCFY96})} model on top of a workflow model. As the name says, RBAC models rely on the concept of \emph{role}, which is different from that of group, as it is a collection of both users and permissions (thus, it acts as an interface between them), rather than a collection of users only~\cite{DBLP:conf/rbac/Sandhu95b}. The main components of an RBAC model are:
\begin{itemize}[noitemsep]
\item $\Users,\Roles,\Perm,\Sessions$ representing the set of users, roles, permissions and sessions, respectively.
\item $\mathit{UA} \subseteq \Users \times \Roles$ and $\PA \subseteq \Roles \times \Perm$ representing many to many user-role and permission-role assignment relations, respectively.
\item $\user : \Sessions \to \Users$ and $\role : \Sessions \to 2^{\Roles}$ representing functions assigning each session to a single user and to a set of roles, respectively.
\item $\RH \subseteq \Roles \times \Roles$ representing a partially ordered role hierarchy relation $\geq$.
\end{itemize}

Moreover, in the classical RBAC model a role $R$ is always \Enabled{} and can
be activated in a session by a user $u$ such that $(u,R) \in \UA$. Thus, to
deal with the lack of constraints on role enabling and disabling,
TRBAC was proposed as a first temporal
extension~\cite{DBLP:journals/tissec/BertinoBF01}. In this model, a user $u$
can activate a role $R$ provided that both he is authorized to do so and the
role is \Enabled{} at the time of the request, i.e., $R \in \ST(t)$, where
$\ST(t)$ is the set of \Enabled{} roles at time $t$. It follows that the
concept of \emph{status} \{\Active{}, \Inactive{}\} of a role is implicitly
augmented by adding \{\Enabled{}, \Disabled{}\}. Note that, if a role is
\Active{} (i.e., there is an associated user playing it in some session),
then it is also \Enabled{}. In general, the vice versa does not hold.

Roles are \Enabled{} and \Disabled{} according to the content of the \emph{role enabling base} (REB) $\mathcal{R}$, which mainly consists of periodic events and triggers. A \emph{periodic event} has the form $(I,P,p\!:\!E)$, where: $I$ is a time interval, $P$ is a periodic expression using calendars~\cite{niezette}, and $p\!:\!E$ is a prioritized event expression where $p$ is a priority and $E$ has the form $\Enable{}\, R$ or $\Disable{}\, R$ for some $R \in \Roles{}$. For example
\begin{displaymath}
(\texttt{[01/01/15,$\infty$], WorkingHours, H:\Enable{} director})
\end{displaymath}
tells the system to enable (with high priority) the role
\texttt{director} from 1 January 2015 onwards, as soon as current time gets
to the starting point of each interval spanned by \texttt{WorkingHours},
which is a periodic expression representing all time instants from 9AM to 1PM
and from 2PM to 6PM of week days (i.e., from Monday to 
Friday).

A \emph{role trigger} has the form $B \to p\!:\!E$  and its meaning is to fire
the periodic event on the right of $\to$ whenever all preconditions on the
left (the body of the trigger consisting of periodic expressions and role
status expressions) become true \cite{DBLP:journals/tissec/BertinoBF01}. A role status expression is either $\Enabled{}\, R$ or $\neg\Enabled{}\, R$. When firing a role trigger these expressions will be evaluated true or false depending on the status of $R$ at that time. For example, the trigger
\begin{displaymath}
\texttt{\Enable{} director $\to$ \Enable{} cashier}
\end{displaymath}
tells the system to enable the role \texttt{cashier} whenever the role
\texttt{director} gets \Enabled.

In this paper, we consider the fragment of TRBAC consisting of
non-conflicting complementary periodic events. For sake of simplicity, we do not consider \emph{non-complementary periodic events} and \emph{conflicting events}. Thus, we assume that each interval spanned by the periodic expression is not influenced by other periodic events.  Moreover, we do not consider \emph{runtime request expressions} (and thus \emph{individual exceptions}) because they allow a security officer to override any execution.
We also do not consider \emph{role triggers} because they may lead to the
previous problems. Thus, under these assumptions, priorities will not influence the behavior of the system.

\section{Access-controlled Workflows}\label{sec:acwf}

We first focus on how a controllable workflow can be executed with respect to a given fragment of TRBAC, where the assumptions made at the end of
Section~\ref{sec:trbacmodels} hold. Then, in Section~\ref{sec:derUsersAuth}
we derive the set of users authorized to execute a time point. In
Section~\ref{sec:caseStudy}, we introduce a running example to discuss a
possible real application. In Section~\ref{sec:scp}, we introduce security
constraints and related propagation rules to enforce security policies at
execution time. In Section~\ref{sec:wfexec}, we discuss how to
execute the access-controlled workflow.


Let us start by supposing that we want to understand whether a controllable
workflow can be executed with respect to the considered fragment of TRBAC.
In such a model, the set $\Perm{} = \{T_1,\dots,T_n\}$ of permissions
consists of the workflow tasks, and the interpretation of the role-permission
assignment relation $(R,T) \in \PA{}$ is ``all users belonging to $R$ are
authorized to execute task $T$''.

As we have already said, roles are \Enabled{} during certain time intervals
and typically \Disabled{} in the complementary. Consequently, since a
workflow task $T$ is represented as a contingent link $A
\xRightarrow[]{[x,y]} C$ where there is no control on the contingent point
$C$, the interval where the associated role is \Enabled{} is supposed to be
at least as large as the maximal duration of the contingent link, i.e., $y$
(otherwise, the workflow would not be consistent with the access control
model). Thus, since the workflow and the access control model do not depend
on each other, we first need to reduce both models to a common
representation to be able to analyze their interplay. To that end, we have
chosen to translate the enabling/di\-sabling intervals of roles belonging to
the access control model into an equivalent STN to be connected to the STNU
representing the workflow model. We proceed as follows:
Section~\ref{sec:pt2stn} introduces a mapping to translate such intervals
into an equivalent STN, and then Section~\ref{sec:configuration} explains how
to connect the resulting STN to the workflow-related STNU.

\subsection{From Periodic Expressions to STNs}\label{sec:pt2stn}
Let $\Cal_i$ be a calendar ($\Hours{},\Days{},\Weeks{},\dots$) and $\Cal_i
\sqsubseteq \Cal_j$ be the sub calendar relation (e.g., $\Days \sqsubseteq
\Weeks$). A \emph{periodic expression} has the form $\sum_{i=1}^n O_i \cdot
\Cal_i \triangleright r \cdot \Cal_d$, where  $O_1 = \mathit{all}$, $O_i \in
2^\mathbb{N} \cup \{\mathit{all}\}$, $\Cal_i \sqsubseteq \Cal_{i-1}$ for
$i=2,\dots,n$, $\Cal_d \sqsubseteq \Cal_n$ and $r \in \mathbb{N}$ \cite{niezette}. The part on the
left of $\triangleright{}$ specifies the set of starting points ($O_i$s) of
the spanned intervals with respect to each calendar ($\Cal_i$) involved, whereas
the part on the right specifies the duration of those intervals in terms of
time units ($r$) in the minimum granularity calendar ($\Cal_d$). For example,
assuming that Monday is the first day of every week, 
\texttt{WorkingHours} can be formalized as follows:
\begin{displaymath}
	\mathit{all}\cdot\Weeks{} + \{1,2,3,4,5\}\cdot\Days{} + \{10,15\}\cdot\Hours{}\,\triangleright\,\{4\}\cdot\Hours{}\,.
\end{displaymath}

Periodic expressions implicitly talk about intervals according to the minimum
granularity chosen (in this case $\Hours{}$). The $x^\mathrm{th}$ hour of the
day starts at time instant $x-1$ and ends at $x$. For instance, the time
instant corresponding to 9AM corresponds to the left bound of the
10$^\mathrm{th}$ hour of the day. Likewise, 1PM corresponds to the right bound
of the 4th hour of the first interval spanned by \texttt{WorkingHours}
starting from the 10$^\mathrm{th}$ (i.e., the 13$^\mathrm{th}$ hour since this
interval contains the 10$^\mathrm{th}$, 11$^\mathrm{th}$, 12$^\mathrm{th}$,
and 13$^\mathrm{th}$ hour of the day).

Every periodic expression can be translated into an equivalent set of simple periodicity constraints. That is, \emph{linear repeating
intervals of integers} along with a \emph{gap constraint} that limits its
applicability as described in~\cite{DBLP:journals/tods/BertinoBFS98} (which
was inspired by~\cite{DBLP:journals/jlp/TomanC98}). Let $P$ be a periodic expression,
$\mathit{Periodicity}(P)$ the number of time units in which $P$ repeats,
$\mathit{Granularity}(P)$ the duration of each spanned interval and
$\mathit{Displacement}(P)$ the set of integers representing the starting points of the spanned
intervals. Then, the \emph{set of equivalent linear repeating intervals of integers} is formally represented by:
\begin{displaymath}
I^P = \{I^P_{n+1,z} \mid 1 \leq y \leq\mathit{Granularity}(P) \ \land 
z \in \mathit{Displacement}(P)\}\,,
\end{displaymath}
where each $I^P_{n+1,z} = [t_1,\dots,t_{\mathit{Granularity}(P)}]$ represents the $(n+1)^\mathrm{th}$ interval of integers spanned by the periodic expression $P$ according to the displacement $z$, and in turn $t_1,\dots,t_{\mathit{Granularity}(P)}$ are generated according to the 
following equation 
representing the class of integers
\begin{displaymath}
t \equiv_{\mathit{Periodicity}(P)} (y + z - 1)\,,
\end{displaymath}
for every $y \in \{1,\dots,\mathit{Granularity}(P)\}$ once we have fixed $z \in \mathit{Displacement}(P)$ and $n \in \mathbb{N}$, where $t \equiv_k c$ denotes the set of integers of the form $c + kn$, ranging from $-\infty$ to $+\infty$ in $\mathbb{Z}$. For each $z \in \mathit{Displacement}(P)$, the set of all start and end points of all intervals spanned by $P$ is computed with respect to $y = 1$ and $y = \mathit{Granularity}(P)$, respectively. 

For example, consider the first complete week of 2015 only (i.e., that
starting on 5 January). \texttt{WorkingHours} translates to $t \equiv_{168} (y
+ z - 1)$ for $y = 1,2,3,4$ and $z=106$, $111$, $130$, $135$, $154$, $159$,
$178$, $183$, $202$, $207$, i.e., from the 10$^\mathrm{th}$ and the
15$^\mathrm{th}$ hour from Monday to Friday of that week for 4 hours.

To get time intervals of real instants, we modify this translation so that it
computes directly such intervals considering only the left and the right
bounds. In other words, given $P$ and an interval $I = \texttt{[begin,end]}$
where $\texttt{begin},\texttt{end}$ are two date expressions identifying a
specific granule according to the minimal granularity adopted\footnote{In this
case, a date expression has the form \texttt{dd/mm/yy:hh} where \texttt{hh}
identifies the (\texttt{hh})$^\mathrm{th}$ hour of \texttt{dd/mm/yy}.} and $0 \leq
\texttt{begin} \leq \texttt{end} < \infty$, the resulting intervals $I^P_{n+1,z}$ can be
computed as $t \equiv_{168} (y + z - 1)$ for $y=0,4$ by fixing each time the value of $n\in\mathbb{N}$ and $z$ as before
(where 168 is the number of hours in a week). Depending on the chosen
$z \in \mathit{Displacement}(P)$, and for $p = \mathit{Periodicity}(P)$ and $g = \mathit{Granularity}(P)$, 
intervals have thus the form
\begin{displaymath}
[pn+z-1,pn+z-1+g]
\end{displaymath}
that instantiated for our \texttt{WorkingHours} becomes
\begin{displaymath}
[168n+z-1,168n+z+3]\,.
\end{displaymath}
For instance, if we restrict the applicability
of $P$ to $I = \texttt{[05/01/15,05/01/15]}$\footnote{When we write  $[\texttt{dd}_1\texttt{/mm}_1\texttt{/yy}_1,\texttt{dd}_2\texttt{/mm}_2\texttt{/yy}_2]$ we mean that $\texttt{hh}_1 = \texttt{01}$ whereas $\texttt{hh}_2 = \texttt{24}$.} (i.e., the first Monday of 2015), then we will obtain:
\begin{align*}
	&\overbrace{[106,107,108,109]}^{\text{05/01/15, (9AM-1PM)}} \cup \overbrace{[111,112,113,114]}^{\text{05/01/15, (2PM-6PM)}}\\ 
	&= 105 \leq t \leq 109 \cup 110 \leq t \leq 114
\end{align*}
where the first line is the computation of linear repeating intervals of
integers, and the second one the conversion in intervals of real time
instants.

Once we have translated $P$ in a finite number of intervals of real values
representing the intervals spanned by $P$ itself, we can generate an
equivalent STN representing them. We will refer to this mapping $\ptToSTN : P
\times I \rightarrow \langle \mathcal{T}, \mathcal{C} \rangle$ as
\emph{periodic time to STN}. Note that for an STN to be generated it is
important that the upper bound of the interval limiting the applicability of
the expression $P$ is $\neq \infty$. Thus, the resulting STN is represented in Figure~\ref{fig:ex:AcSTN}.
\begin{figure}[t]\centering
\begin{tikzpicture}[node distance=50pt]
	\node (Z) {$Z$};
	\node (P1z106S) [above right=of Z,yshift=-25pt] {$P_{1,106}^S$};
	\node (P1z106E) [right=of P1z106S] {$P_{1,106}^E$};
	\node (P1z111S) [below right=of Z,yshift=25pt] {$P_{1,111}^S$};
	\node (P1z111E) [right=of P1z111S] {$P_{1,111}^E$};

	\draw[Link]  (Z) -- node [] {$[105,105]$} (P1z106S);		
	\draw[Link]  (P1z106S) to node {} node [] {$[4,4]$} (P1z106E);		
	\draw[Link]  (Z) -- node [swap] {$[110,110]$} (P1z111S);		
	\draw[Link]  (P1z111S) to node {} node [] {$[4,4]$} (P1z111E);		
\end{tikzpicture}
\caption{$\ptToSTN(\texttt{WorkingHours},\texttt{[05/01/15,05/01/15]})$.}
\label{fig:ex:AcSTN}
\end{figure}

\begin{theorem}\label{thm:pt2stn}
Given any periodic expression $P$ whose applicability is limited by an  interval $I$ whose upper bound is $\neq \infty$, the
mapping $\ptToSTN{}$ returns an equivalent STN that is (i)
consistent and (ii) admits exactly one solution.
\end{theorem}
\begin{CODASPY}
The proof is given in~\cite{CombiViganoZavatteri15}.
\end{CODASPY}
\begin{EXT}
The proof is given in the appendix.
\end{EXT} As a convention, when we write $P_{n+1,z}^{\mathit{sup}}$ we mean the start (if $\mathit{sup}$ is $S$)
or the end (if $\mathit{sup}$ is $E$) point 
of the $(n+1)^\mathrm{th}$ ($n \in \mathbb{N}$) interval spanned by $P$
choosing the displacement $z$ according to the mapping
$\ptToSTN{}$.

Since the TRBAC periodic events we consider are \emph{non-conflicting} and
\emph{complementary}, applying the mapping $\ptToSTN{}$ on the bounded
periodic expressions associated to the periodic events involving only an
``$\Enable{}\,R$'' entails that for each $n$ and $z$, $P_{n+1,z}^S \rightarrow
P_{n+1,z}^E$ represents a time interval (of fixed duration) in which the roles
associated to this expression in the TRBAC role enabling base $\mathcal{R}$
are \Enabled{}. Furthermore, it also turns out that the complementary
intervals have the form $[P_{n+1,z}^E,P_{n+2,z}^S]$, where the bounds
correspond to the real values given by the scheduler $\psi : \mathcal{T} \to
\mathbb{R}$ (which always assigns the same fixed values to these points
depending on $n$ and $z$ and $P$). The next subsection explains how an STN
generated by the mapping $\ptToSTN{}$ restricted to a given upper bound can be
connected to the STNU describing the workflow to check if the workflow itself
can be executed with respect to the given access control model. We point out
that, in general, the mapping $\ptToSTN{}$ returns different STNs depending on
the time window we consider (where a time window is, e.g., the first complete
day of the current year, or the 2nd complete week of the 3rd month of the next
year). Keeping this flexibility allows us to analyze the access control model
in different time windows to leave room for investigating whether or not a
workflow is controllable for all possible STNs generated by $\ptToSTN{}$.

\subsection{Connecting the Access Control Model}\label{sec:configuration}

We are now ready to explain how we can put an access control model on top of a
workflow by connecting the STN describing the access control model to the STNU
describing the workflow. When we connect these two networks we say that the
resulting network, which is still an STNU, is a \emph{configuration}.

Assume a workflow consists of $n$ tasks $T_1,\dots,T_n$ corresponding to the
$n$ contingent links $A_1 \Rightarrow C_1,\dots,A_n \Rightarrow C_n$ in the
workflow STNU. Also, assume the role $R$ is authorized to execute the task
$T$ (i.e., $(R,T) \in \PA{}$) during the time interval $I^P_{n+1,z} =
[P_{n+1,z}^S,P_{n+1,z}^E]$ represented by the
requirement link $P_{n+1,z}^S \xrightarrow[]{[k,k]} P_{n+1,z}^E$, for some
$P$, $n$, $z$ and $k=P_{n+1,z}^E-P_{n+1,z}^S$. Then, to get a configuration
we need to impose that the start of $T$ has to occur \emph{after}
$P_{n+1,z}^S$, whereas the end \emph{before} $P_{n+1,z}^E$.

In other words, role $R$ cannot start $T$ before getting \Enabled{}, and
cannot end it after getting \Disabled{}. A \emph{connection mapping} $\con{} :
\langle \mathcal{T}, \mathcal{C}, \mathcal{L} \rangle \times \langle
\mathcal{T'}, \mathcal{C'} \rangle \to \langle \mathcal{T''}, \mathcal{C''},
\mathcal{L'} \rangle$ is formally depicted in Table~\ref{tbl:connRule}, where
on the left the STN representing the access control (above) and the STNU
representing the workflow (below) are still not connected, whereas on the
right they are.
\begin{CODASPY} Note that due to lack of space the node $Z$ (zero time point)
and the requirement link $Z \xrightarrow[]{[j,j]} P_{n+1,z}^S$ (for some $j =
P_{n+1,z}^S - Z$) have not been shown (but are in~\cite{CombiViganoZavatteri15}).
\end{CODASPY} Also, note that we have added a new
label $\rho$ on the links connecting the STN to STNU. In general, this new
label $\rho = R_1R_2\dots R_n$ consists of a conjunction of roles saying which
roles we want to consider during the interval in which they are \Enabled{} (as
more roles can be \Enabled{} during the same interval).

\begin{lemma}\label{lem:actask}  Given a task represented as a contingent link
$A \xRightarrow[]{[x,y]} C$ connected to a requirement link of access control STN
$P_{n+1,z}^S \xrightarrow[]{[k,k]} P_{n+1,z}^E$ by means of the connection mapping $\con{}$ shown
in Table~\ref{tbl:connRule}, then if $k < y$ the resulting STNU is uncontrollable.
\end{lemma}
\begin{CODASPY}
The proof is given in~\cite{CombiViganoZavatteri15}.
\end{CODASPY}
\begin{EXT}
The proof is given in the appendix.
\end{EXT}

\begin{CODASPY}
\begin{table}[t]
\caption{Connection mapping.}
\label{tbl:connRule}

\begin{tabular}{c}
\toprule

\begin{tikzpicture}[auto,node distance=30pt]
	\node (PnzS) [above right=of Z] {$P_{n+1,z}^S$};
	\node (PnzE) [right=of PnzS] {$P_{n+1,z}^E$};
	\node (A) [below=of PnzS] {$A$};
	\node (C) [below=of PnzE] {$C$};
	
	\draw[Link]  (PnzS) to node {} node [] {$[k,k]$} (PnzE);		
	\draw[ContingentLink]  (A) to node {} node [] {$[x,y]$} (C);

  \node (LEADSTO) [below right=of PnzE,xshift=-20pt,yshift=10pt] {$\leadsto$};

	\node[label={[yshift=-4pt,xshift=-5pt]above:$\{\wf{}\langle\rangle\}$}] (PnzS2) [right=of PnzE,xshift=-4pt]{$P_{n+1,z}^S$};
	\node[label={[yshift=-4pt]above:$\{\wf{}\langle\rangle\}$}] (PnzE2) [right=of PnzS2] {$P_{n+1,z}^E$};
	\node[label={[yshift=4pt,xshift=-3pt]below:$\{u_1\langle\rangle,\dots,u_n\langle\rangle\}$}] (A2) [below=of PnzS2] {$A$};
	\node[label={[yshift=4pt,xshift=3pt]below:$\{u_1\langle\rangle,\dots,u_n\langle\rangle\}$}] (C2) [below=of PnzE2] {$C$};
	
	\draw[Link]  (PnzS2) to node {} node [] {$[k,k]$} (PnzE2);		
	\draw[Link]  (PnzS2) to node {} node [] {$[0,\infty],\rho$} (A2);		
	\draw[Link]  (C2) to node {} node [swap] {$[0,\infty],\rho$} (PnzE2);		
	\draw[ContingentLink]  (A2) to node {} node [] {$[x,y]$} (C2);
\end{tikzpicture}\\
\bottomrule
\end{tabular}
\end{table}
\end{CODASPY}

\begin{EXT}
\begin{table*}[t]\centering
\caption{Connection mapping.}
\label{tbl:connRule}

\begin{tabular}{c}
\toprule

\begin{tikzpicture}[auto,node distance=30pt]
  \node (Z) []{$Z$};
  \node (PnzS) [above right=of Z] {$P_{n+1,z}^S$};
  \node (PnzE) [right=of PnzS] {$P_{n+1,z}^E$};
  \node (A) [below=of PnzS] {$A$};
  \node (C) [below=of PnzE] {$C$};
  
  \draw[Link]  (Z) to node {} node [] {$[j,j]$} (PnzS);   
  \draw[Link]  (PnzS) to node {} node [] {$[k,k]$} (PnzE);    
  \draw[ContingentLink]  (A) to node {} node [] {$[x,y]$} (C);
  \node (Z2) [right=of Z,xshift=120pt]{$Z$};

  \node (LEADSTO) [below right=of PnzE,xshift=-10pt,yshift=10pt] {$\leadsto$};

  \node[label={[yshift=-4pt,xshift=-5pt]above:$\{\wf{}\langle\rangle\}$}] (PnzS2) [above right=of Z2]{$P_{n+1,z}^S$};
  \node[label={[yshift=-4pt]above:$\{\wf{}\langle\rangle\}$}] (PnzE2) [right=of PnzS2] {$P_{n+1,z}^E$};
  \node[label={[yshift=4pt,xshift=-3pt]below:$\{u_1\langle\rangle,\dots,u_n\langle\rangle\}$}] (A2) [below=of PnzS2] {$A$};
  \node[label={[yshift=4pt,xshift=3pt]below:$\{u_1\langle\rangle,\dots,u_n\langle\rangle\}$}] (C2) [below=of PnzE2] {$C$};
  
  \draw[Link]  (Z2) to node {} node [] {$[j,j]$} (PnzS2);   
  \draw[Link]  (PnzS2) to node {} node [] {$[k,k]$} (PnzE2);    
  \draw[Link]  (PnzS2) to node {} node [] {$[0,\infty],\rho$} (A2);   
  \draw[Link]  (C2) to node {} node [swap] {$[0,\infty],\rho$} (PnzE2);   
  \draw[ContingentLink]  (A2) to node {} node [] {$[x,y]$} (C2);
\end{tikzpicture}\\
\bottomrule
\end{tabular}
\end{table*}
\end{EXT}

\subsection{Deriving Authorized Users}\label{sec:derUsersAuth}

Once we have connected the two networks, we are able to derive new
information on which are the users authorized to execute the time points. To
represent this piece of information, we associate to each time point $X \in
\T$ the set $\A(X) \subseteq \Users{}$ of users authorized to execute it as
follows.
\begin{itemize}[noitemsep]
\item For each contingent link $A \Rightarrow C$ representing task $T$, the set of authorized users is $\A(A) = \A(C) = \{u_1\langle c_1 \rangle,\dots,u_n\langle c_n \rangle\}$, where $c_1,\dots c_n$ are security constraints we define in Section~\ref{sec:scp} and $u_1,\dots,u_n$ are users belonging to all roles $R_i$ such that:
\begin{enumerate}[noitemsep]
\item $(R_i,T) \in \PA \land (u_i,R_i) \in \UA$ for $j=1,\dots,n$ (in the TRBAC model), and
\item $R_i$ belongs to $\rho$ specified on the requirement links connecting $P_{n+1,z}^S$ to $A$ and $C$ to $P_{n+1,z}^E$, where $P$ is the associated periodic expression in the periodic event enabling $R_i$ in the REB $\mathcal{R}$. 
\end{enumerate}
\item For each other time point $X$ different from an activation or a contingent point, $\A(X)=\{\wf\langle \rangle\}$ which is a special user we consider to advance the execution of ``internal tasks'' (e.g., branching points). To be more precise we assume that: (i) $\wf \in \Users{}$, (ii) for all $R \in \Roles{}, (\wf,R) \not\in \UA{}$, (iii) for all $X \not\equiv A$ and $\not\equiv C$, $\wf\langle \rangle \in \A(X)$, and (iv) for all $X \equiv A$ or $\equiv C$, $\wf\langle c\rangle \not\in \A(X)$.
\end{itemize}
To conclude this section we extend the form of the classical solution
$\texttt{S} = \{X = t_X,\dots\}$ of a network to the new one $\texttt{S} =
\{(u_i:X=t_X),\dots\}$ for it to take into account who has executed the time
point.
The contingency is modeled by the fact that once a task has started we do not
know when exactly the user will tell the system that he has finished. Of
course, we assume the user to finish within the bounds imposed by the
contingent link, otherwise the system raises an exception to cope with the
situation.

\section{Case study}\label{sec:caseStudy}

Before we discuss \emph{how} to enforce security policies at runtime, we
introduce a running example.

\subsection{Workflow} 
\begin{figure}[t]\centering
\begin{tikzpicture}[node distance=35pt,auto]
	\node[Task,label={above:$R_1$},label={below:$[4,5]$}] (T1) {$T_1$};
	\node[Task,label={above:$R_1$},label={below:$[4,5]$}] (T2) [right=of T1] {$T_2$};
	\node[BranchingPoint,label={below right:$[0,1]$}] (S) [right=of T2] {=};
	
	\node[Task,label={above:$R_2$},label={below:$[1,2]$}] (T3) [above right=of S] {$T_3$};
	\node[Task,label={above:$R_3$},label={below:$[1,3]$}] (T4) [below right=of S] {$T_4$};
	\node[BranchingPoint,label={below right:$[0,1]$}] (J) [below right=of T3] {=};
	
	\draw[Link] (T1) -- node [] {$[1,6]$} (T2);
	\draw[Link] (T2) -- node [] {$[1,1]$}(S);
	\draw[Link] (S) |- node [] {$[0,1]$}(T3);
	\draw[Link] (S) |- node [swap] {$[0,1]$}(T4);
	\draw[Link] (T3) -| node [] {$[1,3]$}(J);
	\draw[Link] (T4) -| node [swap] {$[1,3]$} (J);
\end{tikzpicture}
\caption{Access-controlled workflow. Tasks $T_1$, $T_2$, $T_3$, $T_4$ stand
for \OutwardJourney{}, \ReturnJourney{}, \SystemCheck{} and \SecurityCheck{},
respectively. Roles $R_1$, $R_2$, $R_3$ stand for $\TrainDriver$,
$\SystemEngineer$ and $\SecurityEngineer$, respectively.}
\label{fig:ex:wf}
\end{figure}
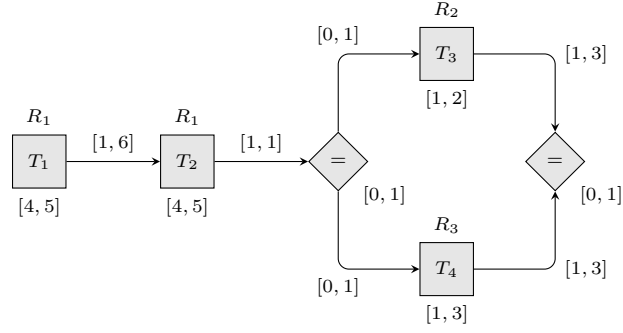

We consider a workflow modeling a round-trip from London to Edinburgh. It
starts with the task \OutwardJourney{} in which the train travels from London
to Edinburgh. The journey takes from 4 to 5 hours to be completed. After the
train has arrived to Edinburgh train station, the \ReturnJourney{} to London
starts within 5 hours since 1 hour after arrival. Once the train has returned,
before the next round trip starts, a \SecurityCheck{} and a \SystemCheck{} are
done in parallel. The first check takes 1 to 2 hours, the second 1 to 3 hours.
Figure~\ref{fig:ex:wf} shows the workflow consisting
of 4 tasks, where we have used the graphical components specified in
Table~\ref{tbl:wf2stnu} (on the left) and decorated each task by a label that
specifies the role authorized to carry it out.

\subsection{Access Control}
The instantiation of the TRBAC is as follows:
\begin{itemize}[noitemsep]
	\item $\Users = \{\Alice,\Bob,\Charlie,\Eve, \Kate\}$
	\item $\Roles = \{\TrainDriver, \SystemEngineer,$ \\
	$ \SecurityEngineer\}$
	\item $\Perm = \{\OutwardJourney{}, \ReturnJourney{}, \\
	\SystemCheck{}, \SecurityCheck{}\}$
	\item $\UA = \{(\Alice, \TrainDriver),(\Bob, \TrainDriver),\\
	(\Charlie, \SystemEngineer), \\
	(\Charlie, \SecurityEngineer),(\Eve, \SecurityEngineer), \\
	(\Kate, \SystemEngineer)\}$
	\item $\PA = \{(\TrainDriver, \OutwardJourney),\\
	 (\TrainDriver, \ReturnJourney),\\(\SystemEngineer, \SystemCheck),\\ (\SecurityEngineer, \SecurityCheck)\}$
\end{itemize}

\SetBoxWidth{9cm} 
\begin{figure}[t]
\begin{empheq}[box={\Garybox[$\mathcal{R}$]}]{align*}
&\texttt{PE}_1 : (\texttt{[01/01/15,$\infty$], $P_1$, \Enable{} $\TrainDriver$})\\
&\texttt{PE}_2 : (\texttt{[01/01/15,$\infty$], $P_2$, \Enable{} $\SystemEngineer$})\\
&\texttt{PE}_3 : (\texttt{[01/01/15,$\infty$], $P_3$, \Enable{} $\SecurityEngineer$})
\end{empheq}
\caption{The Role Enabling Base of the case study.}
\label{fig:ex:REB}
\end{figure}

The role enabling base $\mathcal{R}$ (Figure~\ref{fig:ex:REB}) consists of periodic events only. Each line represents a periodic event as described in Section~\ref{sec:trbacmodels}, 
where the periodic expressions associated to the roles\footnote{We assume
$\mathcal{R}$ also contains the complementary expressions to disable the roles.} are:
\begin{itemize}[noitemsep]
	\item $P_1 = \mathit{all}\cdot\Days{} + \{9\}\cdot\Hours{}\,\triangleright\,\{12\}\cdot\Hours{}$
	\item $P_2 = \mathit{all}\cdot\Days{} + \{16\}\cdot\Hours{}\,\triangleright\,\{9\}\cdot\Hours{}$
	\item $P_3 = \mathit{all}\cdot\Days{} + \{16\}\cdot\Hours{}\,\triangleright\,\{12\}\cdot\Hours{}$
\end{itemize}

In other words, these periodic expressions say that the role $\TrainDriver{}$
is enabled every day from 8AM to 8PM, $\SystemEngineer{}$ every day from 3PM
to 12AM (midnight) and $\SecurityEngineer{}$ from 3PM until 3AM of the day
after. Now consider the time window \texttt{[01/01/15:01,02/01/15:03]}
limiting $P_1,P_2$ and $P_3$. Role $\TrainDriver{}$ is enabled during
$[8,20]$, $\SystemEngineer{}$ during $[15,24]$ and $\SecurityEngineer{}$
during $[15,27]$. Figure~\ref{fig:ex:stnu} shows the resulting configuration.

\subsection{(Temporal) Security Policies}\label{sec:sp}

To conclude the case study section we formalize three security policies that
are supposed to hold in this context.

\begin{securitypolicy}\label{sp:1}
\emph{A user who starts a task ends it too.}
\end{securitypolicy}

\begin{securitypolicy}\label{sp:2}
\emph{A user is allowed to execute no more than one task at a time}.
\end{securitypolicy}

\begin{securitypolicy}\label{sp:3}
\emph{If the train driver from Edinburgh to London is the same as the one who
drove the train from London to Edinburgh, he must rest at least 2 hours
before driving again.}
\end{securitypolicy}

We assume that policies~\ref{sp:1}, \ref{sp:2} and~\ref{sp:3} hold for our
case study.

Without security constraint propagation rules, it is rather easy to note that, 
if every time we start or end a task we are free to choose any authorized user
among those contained in that set, then our example violates all the previous
security policies. Therefore, the next section provides a language to define
security constraints along with a set of security constraint propagation rules
to propagate them when executing.

\section{Propagation of Security Const\-raints}\label{sec:scp}
In Section~\ref{sec:derUsersAuth}, we discussed how to
compute the set of authorized users for a time point once we have obtained a
configuration (Section~\ref{sec:configuration}). We now proceed by giving its
canonical form as further contribution.

\begin{definition}\em
For each time point $X$ belonging to a configuration, the set of authorized
users has the \emph{canonical form} $\A(X) = \{u_1\langle c_1 \rangle,
u_2\langle \mathit{c}_2 \rangle,\dots,u_n\langle c_n \rangle\}$, where
$u_1,\dots,u_n$ are the authorized users and $c_1,\dots,c_n$ are (temporal)
security constraints defined according to the grammar
\begin{align*}
c \ &::= \ t_k,\Box \mid X + k,\Box \\
\Box \ &::= \ >\,\mid\,<\,\mid\,\geq\,\mid\,\leq\,\mid\,=\,\mid\,\neq\,
\end{align*}
where $t_k,k \in \mathbb{R}^{\geq0}$. $c = t_k,\Box$ is a \emph{Type-1 security
constraint}, whereas $c = X + k,\Box$ is \emph{Type-2}.
\end{definition}
Every Type-2 security constraint $c = X+k,\Box$ is reducible to a Type-1 by
substituting $X$ with a real value taken from its range, plus $k$ (if $k \neq
0$). As an example, let $c = X + 1, \leq$. Once the value of
$X$ becomes known, say the scheduler executes it at time $3$ ($\psi(X)=3$), we
substitute it for $X$ also adding the constant $k=1$. After that, the Type-2
security constraint reduces to the Type-1 $c' = 4, \leq$.
\begin{displaymath}
	\underbrace{c = X + 1,\leq}_{t < 3}\quad \overset{\psi(X)=3}{\leadsto}\quad \underbrace{c' = 4,\leq}_{t\geq 3}
\end{displaymath}
The main idea behind a security constraint is that of blocking the associated
user in order to prevent him from executing some time point if a security
policy is violated.

\begin{definition}\em
We define \emph{interpretation} of security constraints with respect to current time $t$ as follows:
\begin{multicols}{2}
\begin{enumerate}[noitemsep]
	\item $t \models t_k, >$ iff $t > t_k$
	\item $t \models t_k, <$ iff $t < t_k$
	\item $t \models t_k, \geq$ iff $t \geq t_k$
	\item $t \models t_k, \leq$ iff $t \leq t_k$
	\item $t \models t_k, =$ iff $t = t_k$
	\item $t \models t_k, \neq$ iff $t \neq t_k$
	\item $t \not\models X + k, >$  
	\item $t \models X + k, <$ 
	\item $t \not\models X + k, \geq$
	\item $t \models X + k, \leq$
	\item $t \not\models X + k, =$
	\item $t \models X + k, \neq$
\end{enumerate}
\end{multicols}
A user $u$ is \emph{blocked} for the time point $X$ if $u \langle c \rangle \in \A(X)$, and current
time $t \models c$.
\end{definition}

It is clear from the context that 1--6 regard Type-1 security constraints,
whereas 7--12 regard the Type-2. The interpretation of the first group with
respect to the chosen $\Box{}$ operator substantially states that those
constraints will be true if current time $t$ is greater than the value
specified (1), less than it (2), greater than or equal to it (3), less than
or equal to it (4), equal to it (5), or different from it (6). Instead, that
of the second group (Type-2) is a little bit subtle since the value of time
point specified in it is still unknown.

The interpretation with respect to the chosen $\Box{}$ operator states that,
since $X$ will be executed in the future and is thus yet unknown, the Type-2
constraint is interpreted false (7), true (8), false (9), true (10), false
(11) and true (12). When $X$ executes the Type-2 constraint is
reduced to a Type-1 and interpreted accordingly.

As an example, consider $c = X + 3, >$. The current time satisfies this
constraint iff it is greater than $X+3$, where the value of $X$ is still
unknown. Therefore, at current time $t$ this constraint is false.

Furthermore, when $X$ executes, this constraint remains false for $3$ other
time units. Now consider the complementary case $c = X + 3, \leq$. The current
time satisfies $c$ iff it is less than or equal to $X+3$, where the value of
$X$ is still unknown. Even if $X$ is still unknown, this constraint at time
$t$ is trivially true since current time is of course less than (or equal to)
some other value in the future plus some \emph{positive} constant.
Furthermore, when $X$ executes, this constraint remains true for $3$ other
time units. Similar explanations apply to the other Type-2 constraints.

Security constraint propagation rules say how security constraints propagate
when executing.
\begin{definition}\em
A \emph{security constraint propagation rule (SC\-PR)} is a 4-tuple of the
form: 
\begin{displaymath}
  \langle X, \langle c \rangle, \Y, \Diamond \rangle\,
\end{displaymath} 
where $X$ is a time point, $c$ a
security constraint, $\Y$ a set of time points, and $\Diamond$ is either $=$
or $\neq$.
\end{definition}

The semantics of an SCPR says that when $X$ is executed, the security
constraint $c$ has to be set to all users in $\A(Y)$ ($Y \in \Y{}$) equal to
(if $\Diamond$ is $=$) or different from (if $\Diamond$ is $\neq$) the user
who executed $X$.

\begin{figure*}[t]\centering
\scalebox{0.85}{\begin{tikzpicture}[node distance=40pt,auto]

	\node[label={[xshift=-5pt]right:$\{\wf\}$}] (Z) {$Z$};
	\node[label={[yshift=4pt]below:$\{u_1\langle\rangle,u_2\langle\rangle\}$}] (A1) [right=of Z] {$A_1$};
	\node[label={[yshift=4pt]below:$\{u_1\langle\rangle,u_2\langle\rangle\}$}] (C1) [right=of A1]{$C_1$};
	\node[label={[yshift=4pt]below:$\{u_1\langle\rangle,u_2\langle\rangle\}$}] (A2) [right=of C1,xshift=-10pt] {$A_2$};
	\node[label={[yshift=4pt]below:$\{u_1\langle\rangle,u_2\langle\rangle\}$}] (C2) [right=of A2] {$C_2$};
	\node[label={[yshift=4pt]below:$\{\wf\langle\rangle\}$}] (BS) [right=of C2,xshift=-20pt] {$B^S$};
	\node[label={[yshift=4pt,xshift=-3pt]below:$\{\wf\langle\rangle\}$}] (BE) [right=of BS,xshift=-20pt] {$B^E$};
	\node[label={[xshift=5pt,yshift=5pt]left:$\{u_3\langle\rangle,u_5\langle\rangle\}$}] (A3) [above right=of BE,yshift=-10pt] {$A_3$};
	\node[label={[xshift=-5pt,yshift=5pt]right:$\{u_3\langle\rangle,u_5\langle\rangle\}$}] (C3) [right=of A3] {$C_3$};
	\node[label={[xshift=5pt,yshift=-5pt]left:$\{u_3\langle\rangle,u_4\langle\rangle\}$}] (A4) [below right=of BE,yshift=10pt] {$A_4$};
	\node[label={[xshift=-5pt,yshift=-5pt]right:$\{u_3\langle\rangle,u_4\langle\rangle\}$}] (C4) [right=of A4] {$C_4$};
	\node[label={[yshift=4pt,xshift=3pt]below:$\{\wf\langle\rangle\}$}] (ES) [right=of BE,xshift=90pt] {$E^S$};
	\node[label={[yshift=4pt]below:$\{\wf\langle\rangle\}$}] (EE) [right=of ES,xshift=-20pt] {$E^E$};

	\draw[ContingentLink] (A1) -- node {$[4,5]$} (C1);
	\draw[ContingentLink] (A2) -- node {$[4,5]$} (C2);
	\draw[ContingentLink] (A3) -- node {$[1,2]$} (C3);
	\draw[ContingentLink] (A4) -- node {$[1,3]$} (C4);
	\draw[Link] (C1) -- node {$[1,6]$} (A2);
	\draw[Link] (C2) -- node {$[1,1]$} (BS);	
	\draw[Link] (BS) -- node {$[0,1]$} (BE);	
	\draw[Link] (BE) -- node []{$[0,1]$} (A3);	
	\draw[Link] (BE) -- node [] {$[0,1]$} (A4);
	\draw[Link] (C3) -- node {$[1,3]$} (ES);
	\draw[Link] (C4) -- node [] {$[1,3]$} (ES);
	\draw[Link] (ES) -- node {$[0,1]$} (EE);	

	\node[label={[xshift=5pt]left:$\{\wf\langle\rangle\}$}] (PR1S) [above=of A1,xshift=-15pt,yshift=-15pt] {$P_1^S$};
	\node[label={[yshift=-4pt]above:$\{\wf\langle\rangle\}$}] (PR1E) [above=of C1,xshift=-15pt,yshift=-15pt] {$P_1^E$};
	\node[label={[yshift=-4pt]above:$\{\wf\langle\rangle\}$}] (PR2S) [above=of A3,yshift=-15pt] {$P_2^S$};
	\node[label={[yshift=-4pt]above:$\{\wf\langle\rangle\}$}] (PR2E) [above=of C3,yshift=-15pt] {$P_2^E$};
	\node[label={[yshift=4pt]below:$\{\wf\langle\rangle\}$}] (PR3S) [below=of A4,yshift=15pt] {$P_3^S$};
	\node[label={[yshift=4pt]below:$\{\wf\langle\rangle\}$}] (PR3E) [below=of C4,yshift=15pt] {$P_3^E$};
	\path[Link] (Z) edge [] node [auto,sloped, anchor=center, above] {$[8,8]$} (PR1S);

	\draw[Link] (Z) |- node [near end] {$[15,15]$} (PR2S);	
	\draw[Link] (Z) |- node[swap,near end] {$[15,15]$} (PR3S);	
	\draw[Link] (PR1S) -- node {$[12,12]$} (PR1E);
	\draw[Link] (PR2S) -- node {$[9,9]$} (PR2E);
	\draw[Link] (PR3S) -- node {$[12,12]$} (PR3E);
	\draw[Link]  (PR1S) to node {} node [] {$[0,\infty],R_1$} (A1);		
	\draw[Link]  (C1) to node {} node [swap] {$[0,\infty],R_1$} (PR1E);		
	\path[line] (PR1S) -- ++ (0pt,18pt) node[] {}  -| node [near start] {$[0,\infty],R_1$} (A2);
	\draw[Link]  (C2) |- node [swap] {$[0,\infty],R_1$} (PR1E);		
	\draw[Link]  (PR2S) to node {} node [] {$[0,\infty],R_2$} (A3);		
	\draw[Link]  (C3) to node {} node [swap] {$[0,\infty],R_2$} (PR2E);		
	\draw[Link]  (PR3S) to node {} node [swap] {$[0,\infty],R_3$} (A4);		
	\draw[Link]  (C4) to node {} node [] {$[0,\infty],R_3$} (PR3E);		
\end{tikzpicture}}
\caption{STNU equivalent to the access-controlled workflow depicted in Figure~\ref{fig:ex:wf}. 
Users $u_1$, $u_2$, $u_3$, $u_4$, $u_5$ represent $\Alice$, $\Bob$, $\Charlie$, $\Eve$, and $\Kate$, respectively.}
\label{fig:ex:stnu}
\end{figure*}
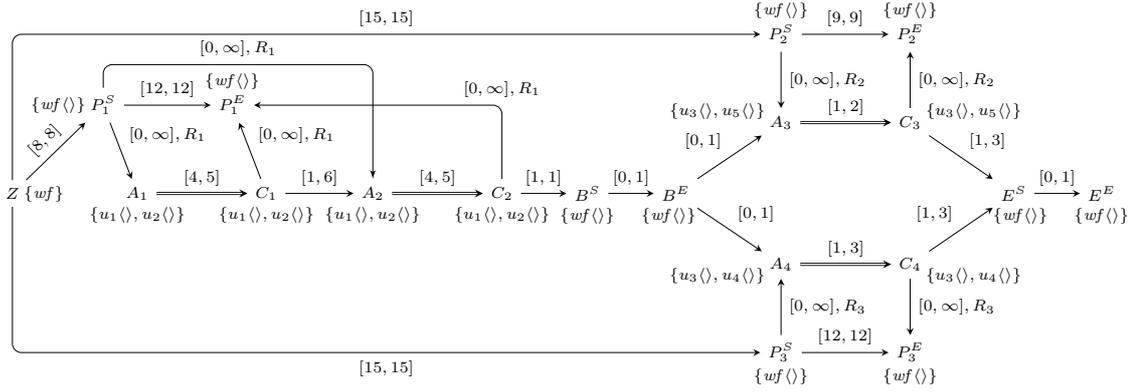

\subsection{From Security Policies to SCPRs}

We can use SCPRs as a means to embed the (temporal) security policies we want
to hold. We exemplify this at hand of the three security policies defined in
Section~\ref{sec:sp} by sketching a few constructs of an high-level language
to define security policies for the workflow we are currently designing. This
language allows Security Officers to specify security policies in an easier
way without even mentioning time points. Then, an intermediate step is that of
generating a set $S$ of SCPRs starting from the constructs of this language for the
system to be able to do a safeness check first and propagate SCs while
executing.

We know that by means of the mapping $\ptToSTN{}$ each task $T$ is represented
as a contingent link describing its start and end point. Thus we need
constructs such as $\texttt{start}(T)$ end $\texttt{end}(T)$ to model these
aspects.

To express conditions on who did what, we envision to have primitives like
$\texttt{hasExecuted}(u,T)$, $\texttt{hasStarted}(u,T)$,
$\texttt{hasEnded}(u,T)$ as binary predicates modeling the fact that user $u$
has executed/started/ended task $T$.

Instead, to model who is not allowed to start or end a task we envision to
have primitives $\texttt{cannotStart}(u,T)$, $\texttt{cannotEnd}(u,T)$ as well
as clauses such as $u_1 = u_2$ and $u_1 \neq u_2$ (resp. $T_1 = T_2$ and $T_1
\neq T_2$) to intend the same or a different user (resp. task).

This language also needs to quantify over the sets of authorized users and
tasks to formalize properties such as \texttt{for all ... [in ...]} other than
conditional blocks such as \texttt{if ... then ... [else ...]}. Note
that since we are considering structured workflows, we are able to formalize
statements such as ``for all tasks in the first parallel block''.

Last but not least, we need ``temporal constructs'' such as \texttt{before} $k$
\texttt{after} $\texttt{end}(T)$ to model security properties such as Temporal Separation of Duties (TSoD).

\emph{Security Policy~\ref{sp:1}} requires that an authorized user who starts a task ends it too. 
In the high-level language we expect to formalize it this way:
\begin{align*}
\texttt{for }& \texttt{all } T \texttt{ if } \texttt{hasStarted}(u,T) \texttt{ then }\\
						&\texttt{for all } u' \texttt{ in } \A(T)\texttt{ } \\
						& \texttt{if } u' \neq u \texttt{ then } \texttt{cannotEnd}(u',T)
\end{align*}
This rule is translated in $n$ SCPRs having the form:
\begin{align*} 
&r_i : \langle A_i, \langle C_i,\leq \rangle, C_i, \neq \rangle 
\end{align*}
where $n$ is the number of tasks which the workflow consists of.
In our case study is translated to 4 rules (since the workflow of Figure~\ref{fig:ex:wf} consists of 4 tasks):
\begin{align*} 
&r_1 : \langle A_1, \langle C_1,\leq \rangle, C_1,
\neq \rangle & r_3 : \langle A_3, \langle C_3,\leq \rangle, C_3, \neq
\rangle\\ &r_2 : \langle A_2, \langle C_2,\leq \rangle, C_2, \neq \rangle &
r_4 : \langle A_4, \langle C_4,\leq \rangle, C_4, \neq \rangle 
\end{align*}
That is, every time an authorized user $u$ executes an activation point $A$,
the system sets the constraint $C, \leq$ to all users $u' \in \A(C)$ where $u'
\neq u$. Those users will be blocked until $t \models C,\leq$ (i.e., until $C$
is executed).

\emph{Security Policy~\ref{sp:2}} requires that an authorized user is allowed to execute one task at a time.
\begin{align*}
\texttt{for }&\texttt{all } T \texttt{ in }\texttt{ParallelBlock} \texttt{ if } \texttt{hasStarted}(u,T) \texttt{ then}\\
						&\texttt{for all } T' \texttt{ in ParallelBlock} \texttt{ if } T' \neq T \texttt{ then }\\
            &\qquad\texttt{cannotStart$(u,T')$ before end$(T)$}
\end{align*}
This rule is translated in $n$ SCPRs having the form:
\begin{align*} 
r_i : \langle A_i, \langle C_i,\leq \rangle, \{A_j\}, = \rangle
\end{align*}
where $i \neq j$ and $n$ is the number of tasks in the considered parallel block (because we model all possible cases).
In our case study is translated to:
\begin{align*}
& r_5 : \langle A_3, \langle C_3,\leq \rangle, \{A_4\}, = \rangle & r_6 : \langle A_4, \langle C_4,\leq \rangle, \{A_3\}, = \rangle
\end{align*}
In the case study introduced in Section~\ref{sec:caseStudy} there is only one
parallel block which consists of tasks $T_3,T_4$ (Figure~\ref{fig:ex:wf}) or
equivalently contingent links $A_3 \Rightarrow C_3,A_4 \Rightarrow C_4$
(Figure~\ref{fig:ex:stnu}). Therefore, if a user can execute both tasks (where
the execution order of $T_3,T_4$ is not well defined), we must prevent him
from executing the other until the current is not finished. Relying on $r_1$
it is enough to set the constraint (i.e., to block the user) only on the
activation points $A_3,A_4$ depending on which task executes first.
	
\emph{Security Policy~\ref{sp:3}} requires that if the same authorized user executes $T_1$ and $T_2$, then between the end of $T_1$ and the start of $T_2$ at least 2 hours have to elapse. 
\begin{align*}
\texttt{for all }& u \texttt{ in } \A(T_1) \texttt{ if } \texttt{hasStarted}(u,T_1) \texttt{ then }\\
						&\texttt{cannotStart}(u,T_2) \texttt{ before } 2 \texttt{ after } \texttt{end}(T_1) 
\end{align*}
This rule is translated in a single SCPR having the form:
\begin{align*} 
r_i : \langle C_i, \langle C_i+2,\leq \rangle, \{A_j\}, = \rangle
\end{align*}
where $i \neq j$.
In our case study is translated to:
\begin{align*}
&r_7 : \langle C_1, \langle C_1+2,\leq \rangle, \{A_2\}, = \rangle
\end{align*}
The same reason of $r_5,r_6$ (i.e., \emph{relying on $r_1$}) applies here to avoid writing a redundant $C_2$ in $r_7$'s $\Y{}$. 

\emph{Separation of Duties (SoD)} requires that the same user ought not be
able to carry out two sensitive tasks in the same execution. \emph{Temporal
Separation of Duties (TSoD)} is an extension of SoD that allows the same user
to do so, but \emph{only if} a further temporal constraints is satisfied. In
our example, the train driver can bring the train back to London as long as
he has rested at least 2 hours (the further temporal constraint).
Furthermore, imposing this constraint at workflow level (e.g., tightening the
STNU by means of a requirement link $C_1 \xrightarrow[]{[2+\epsilon,\infty]}
A_2$, for some $\epsilon > 0$) would be wrong. Indeed, it would prevent
\emph{all} train drivers different from the one who drove the train during
the outward journey from driving the train in the return journey \emph{as
soon as possible} since the arrival at Edinburgh station. That is, exactly
after 1 hour considering that we have the constraint $C_1
\xrightarrow[]{[1,6]} A_2$ (Figure~\ref{fig:ex:stnu}).

\subsection{Safeness of a set of SCPRs}

We are left to specify the notion of \emph{safeness} for a set $S$ of SCPRs.
We recall that an STNU is executed
incrementally~\cite{DBLP:conf/aaai/MorrisM00}. To do that, we must define (i) \emph{conflicting rules}, and (ii) an \emph{algorithm} to check if a set $S$ of SCPRs does not contain any pair of conflicting rules. We proceed by providing the definition of conflicting rules and then a safeness check algorithm. 
 
\begin{definition}\label{def:confrule}\em
Two SCPRs $r_1 = \langle X_1, c_1, \Y_1, \Diamond_1 \rangle$ and $r_2 = \langle X_2,c_2,$ $\Y_2, \Diamond_2 \rangle$ are in \emph{conflict} iff the following four conditions hold all together: (1) $X_1 = X_2$, (2) $c_1 \neq c_2$, (3) $\Y_1 \cap \Y_2 \neq \emptyset$, and (4) $\Diamond_1 = \Diamond_2$.
\end{definition}
As an example, the following two SCPRs are conflicting:
\begin{align*}
  & r_1:\langle X, \langle Y+2,\leq \rangle, \{Z,V\}, \neq \rangle & r_2:\langle X, \langle Y,> \rangle, \{W,Z\}, \neq \rangle
\end{align*}
Rule $r_1$ says that when $X$ has been executed by $u \in \A(X)$ the security constraint
$Y+2,\leq$ has to be set for all users different from $u$
belonging to the sets $\A(Z)$ and $\A(V)$, whereas $r_2$ says that the complementary constraint ($Y,>$) has to be set, again, for all
users different from $u$ belonging to $\A(W)$ and $\A(Z)$. These two rules are
conflicting since the four conditions hold all together: (1) $X=X$, (2)
$Y+2,\leq\,\, \neq Y,>$, (3) $\{Z,V\} \cap
\{W,Z\} = \{Z\} \neq \emptyset$, and (4) both $\Diamond$ are $\neq$. In other
words, they are trying to set \emph{at the same time} a different constraint for the same users
belonging to $\A(Z)$.

\begin{algorithm}[t]\small
  \caption{A safeness-checker for a set $S$ of SCPRs}
  \begin{algorithmic}[1]
    \Procedure{$\mathit{SafenessChecker}$}{$S$} 
	\For{\textbf{all} $r_1,r_2 \in S$}
		\If{$r_1 \neq r_2$  and $r_1$ and $r_2$ are conflicting}
			\State \Return{$\mathit{false}$}
		\EndIf{}
	\EndFor
	\State \Return{$\mathit{true}$}
	\EndProcedure
  \end{algorithmic}
  \label{alg:safenessChecker}
\end{algorithm}

It is quite easy now to see that the set $S=\{r_1,\dots,r_7\}$ containing the
seven SCPRs translating the three security policies of our case study is
\emph{safe}, since there does not exist any pair of conflicting rules. A
first brute-force procedure to check the safeness of a set $S =
\{r_1,\dots,r_n\}$ of SCPRs is given in Algorithm~\ref{alg:safenessChecker},
which takes as input $S$ and tests the
four conditions of Definition~\ref{def:confrule} for all different pairs of
rules $r_i,r_j \in S$. The algorithm runs exactly in $\Theta(|S|^2)$.

\section{Workflow Execution}\label{sec:wfexec} 

In order to propagate security constraints when the workflow is being
executed we are left to do one thing: to specify how SCPRs are taken into
account at runtime. To do so, we extend the STNU execution algorithm\begin{EXT}\footnote{The classic algorithm 
starts by inserting all control points 
in the set \texttt{A}. Then, while \texttt{A}
is not empty, it executes incrementally the enabled time points removing them
from it and building the solution \texttt{S}. A time point is \emph{live} if
current time lies between its upper and lower bounds, and \emph{enabled} if it
is \emph{live} and all time points to be executed before it have already been
executed~\cite{DBLP:conf/aaai/MorrisM00}. In the original version
\cite{DBLP:conf/aaai/MorrisM00} the solution has the form $\texttt{S} = \{X = t_X,\dots\}$.}\end{EXT} in~\cite{DBLP:conf/aaai/MorrisM00} for
it to take as input also a (safe) set of SCPRs that shall be evaluated each
time a time point is executed. If the executed time point matches the guard of
some rules, then the constraints in them have to be set (thus propagated)
before the execution continues according to all we have said so far
(Algorithm~\ref{alg:executor}).

\begin{algorithm}[t]\small
  \caption{A configuration execution algorithm}
  \begin{algorithmic}[1]
    \Procedure{$\mathit{Executor}$}{$S$} 
    \State{\texttt{A} = set of all control points of the network}
	\State{$t=0$}
	\While{$\texttt{A} \neq \emptyset$}
	\State{Wait for some time point live and enabled $X\in\texttt{A}$}
	\State{Arbitrary pick a live and enable time point $X\in \texttt{A}$}
	\State{Arbitrary pick $u\langle c \rangle \in \A(X)$ such that $t \not\models c$}
	\State{$\texttt{S} = \texttt{S} \cup (u:X=t)$}
	\State{Propagate all $\langle X_i,c_i,\Y_i,\Diamond_i \rangle \in S$ s.t. $X = X_i$}
	\State{$\!\!$Advance $t$ propagating all temporal constraints}
	\EndWhile{}
	\EndProcedure
  \end{algorithmic}
  \label{alg:executor}
\end{algorithm}

%
Now it is time to be more concrete by writing down how the state of the system
evolves during the execution. Of course, there are infinite different ways of
executing  a workflow since time is dense; consequently, each range $[x,y]$
where $x \neq y$ (e.g., those belonging to contingent links) consists of
infinite points. For instance, in our case study we have chosen \emph{a
possible execution} with the only purpose of showing how security constraints
propagate. The execution is given in Table~\ref{tbl:caseStudy} (in the
appendix), where cells in bold-face point out the security constraint(s) being
applied.

\section{Related Work}\label{sec:RW}

Related work on defining, validating, and enforcing access control policies in workflow contexts can be grouped in four main areas: (i) access control and
workflow models, (ii) authorization constraints, (iii) planning, and (iv)
run-time execution.

RBAC models~\cite{DBLP:journals/computer/SandhuCFY96} are the default choice for many organizations that need to balance security with flexibility.
Classical RBAC models are however unable to deal with security policies at
user level, such as separation or binding of duties. 
 
In~\cite{DBLP:journals/tissec/BertinoFA99}, Bertino et al. give a language for
defining authorization constraints on role and user assignment to tasks in a
workflow. They also provide algorithms for constraint consistency check and
task assignment. This proposal assumes the workflow to verify a total order on tasks (i.e., no parallel tasks are allowed). Furthermore, temporal
constraints are not investigated.

The Temporal Authorization Base model described
in~\cite{DBLP:journals/jcs/BertinoBFS00} is able to enforce authorization constraints in heterogeneous distributed systems. It allows users to assign periodic authorizations to other users on sets of objects. This model is quite expressive. In order to use it in a workflow context, we
conjecture that it would be required to restrict access modes to
\texttt{execute} and constrain objects to be \texttt{tasks}. However, a
thorough investigation is needed, as in our context we also need to deal with the temporal aspects related to  workflows.


A number of proposals (Wang and Li in~\cite{DBLP:journals/tissec/WangL10} and Crampton et al. in~\cite{DBLP:journals/jair/CohenCGGJ14,DBLP:conf/sacmat/Crampton05, DBLP:conf/iwpec/CramptonGGJ15,DBLP:journals/tissec/CramptonGY13, DBLP:journals/sttt/CramptonHK14}, to name a few) have addressed the workflow satisfiability problem (WSP) and the resiliency problem. WSP is the problem of assigning tasks to users so that the execution of the access controlled workflow is guaranteed to reach the end, when dealing with authorization constraints that might prevent some user from doing some action in the future.
In general, solving WSP requires exponential time, but some of the cited approaches proved (by using parameterized complexity and kernelization) that for some workflow instances (those where the number of users equals the number of tasks) the WSP can be solved in polynomial time. In our work, we have not dealt with WSP yet, but only with the satisfaction of temporal constraints. As for current and future work, we are trying to extend our approach to deal with WSP, by refining the classical STNU DC-check to take into account authorization constraints as well. If an STNU passes this check, then we can also generate an execution strategy to execute the workflow (planning phase), being guaranteed to always
get to the end by satisfying all constraints. 

On the other hand, the \textit{resiliency problem} faces the issue of how to deal with the execution of plans if some authorized users become unavailable, when the workflow is being executed. To the best of our knowledge, no solution to the WSP and to the resiliency problems using temporal networks has been proposed so far and the related temporal aspects of such problems deserve further attention.

Finally, in~\cite{DBLP:journals/jwsr/PaciBC08}, Paci et al. proposed proposed \emph{RBAC-WS-BPEL}, a role-based access control model for the web services business process execution language \emph{WS-BPEL}. The language is based on XML and does not deal with temporal constraints, which makes it unsuitable for our context.


\section{Conclusions and Future work}\label{sec:CFW}

In this paper we focused on the issue of managing in a seamless way role-based access control mechanisms in temporal workflows, where temporal aspects are both in the access control model and in the workflow model. We based our approach on temporal constraint networks.
To the best of our knowledge, all of the existing approaches are
unsuitable to deal with a workflow having both temporal constraints related to the execution control model and in the access control model.  

We have provided mappings to translate a workflow into an equivalent STNU and a time window of TRBAC into an equivalent STN to be connected to STNU
describing the workflow. This allowed us to answer the question about whether or not the resulting configuration is executable. Then, we have derived for each time point the set of authorized users and defined security constraints along with their propagation rules in order to set, update and propagate security constraints also discussing safeness and execution issues.

We view this paper as a first step of a more general research work, where we will consider and face the following research directions: considering
\emph{conditional STNUs} \cite{HunsbergerPC12} in order to
augment expressiveness to model different workflow paths; allowing the \emph{combination of temporal security constraints} such as $u \langle c_1 \land c_2 \rangle$ or $u \langle \neg c \rangle$ to express security policies such as ``(not) during''; defining a \emph{high-level security policy language}; and considering the \emph{temporal workflow
satisfiability problem} (TWSP) (i.e., \emph{controllability with respect to
security}), to do an automated validation of security policies with respect to different possible temporal configurations and user assignments.

\appendix



\begin{EXT}
\section{Example of execution of the case study}
\end{EXT}

Train TR2015 from London to Edinburgh departs at 08:00 AM (from platform
1). The system ($\wf{}$) starts the workflow at 12AM of the 1st of January
2015. That is, it executes the zero time point $Z$ without propagating any
security constraint. The system also executes at 8 o'clock $P_1^S$ enabling
$R_1$.

At 8 o'clock Bob ($u_2$), who is a train driver, starts the outward journey
from London to Edinburgh. Rule $r_1$ (enforcing Security Policy~\ref{sp:1})
constrains $\Bob{}$ to be the only user to end the task (i.e. to execute
$C_1$) between 12PM and 1PM by applying the security constraint $C_1, \leq$
for all users apart from him in $\A(C_1)$ (row 1). Suppose the outward journey
takes 4 hours. Thus, $\Bob{}$ tells the system the train has arrived at
Edinburgh station at midday (row 2). Furthermore, rule $r_7$ (enforcing
Security Policy~\ref{sp:3}) prevents $\Bob{}$ from starting the return journey
before current time does not pass 2 o'clock by applying the security
constraint $14, \leq$\footnote{Instead of applying $C_1+2,\leq$ (as formalized
in $r_1$), the system applies directly $14, \leq$ since the value of $C_1$
($12$) is known.} for him in $\A(A_2)$ (again, row 2). Meanwhile at 3 o'clock
the system starts both $P_2^S$ and $P_3^S$ enabling $R_2$ and $R_3$,
respectively without applying any security constraint.

Now let us say that Bob starts the return journey at 3 o'clock. Rule $r_1$
constrains him to finish the journey between 7PM and 8PM. That is, $C_2, \leq$
is applied for all users apart from him in $\A(C_2)$ (row 3). Assume this time
the return journey takes its maximal duration (5 hours). Since there is no
security policy which says something upon the arrival of the return journey
there is not any rule $r$ whose guard is $C_2$ either (row 4). The system also
disables $R_1$ by executing $P_1^E$ at the same time.

Assume now the system decides the duration of the branch block starting the
parallel (which can be viewed as an internal task) is instantaneous and starts
exactly after 1 hour since the train has got back to London. That is, the
system executes $B^S$ and $B^E$ at 9 o'clock without applying any security
constraint.

Suppose now that Charlie ($u_3$) starts the system check at 9PM. Rule $r_5$
(enforcing Security Policy~\ref{sp:2}) prevents him from executing the
security check until he has finished the current task by applying $C_3, \leq$
for him in $\A(A_4)$ (row 5). The motivation is that Charlie is both a System
and a Security Engineer, thereby he is authorized to execute both tasks.
Furthermore, $r_3$ fires too by constraining Charlie to be the only one
authorized to end the task applying $\langle C_3, \leq \rangle$ for all users
apart from him in $\A(C_3)$ (again, row 5).

Assume now that while system check is being executed, Eve ($u_4$) starts the
security check at 10PM. Rule $r_6$ does the same of $r_5$ but with respect to
Eve and task $\SystemCheck{}$. However, since Eve is not a System Engineer
(consequently $u_4 \not\in \A(A_3)$) this rules has no effect on the state of
the system (row 6). Instead, $r_4$ applies as usual by setting $C_4, \leq$ for
all users apart from Eve in $\A(C_4)$ (again, row 6). Now suppose Charlie and
Eve terminate the tasks they are executing at 11PM and at 1AM (of the day
after), respectively (rows 7 and 8). What happens next is that no security
constraint is applied (because there are no rules whose guards contain $C_3$
or $C_4$) and the system disables $R_3$ at midnight (by executing $P_2^E$) and
$R_4$ at 3AM of the day after (by executing $P_3^E$).

Finally, as for the branch block, the system decides the duration of the join
block is instantaneous and starts after 1 hour since the last task
($\SecurityCheck{}$ in this strategy) has terminated, i.e., the system
executes $E^S$ and $E^E$ at 1AM of the day after without applying any security
constraint. \\

\begin{table*}[t]
\caption{Example of execution of the case study. $\A(A_i)$ (resp., $\A(C_i)$ is the set of users authorized to start (resp., end) task $T_i$. The first column shows when which user has executed which time point. 
}
\label{tbl:caseStudy}
\scalebox{0.55}{\begin{tabular}{*{9}{>{$}c<{$}}}
\toprule

\texttt{Executed TP} & \A(A_1) & \A(C_1) & \A(A_2) & \A(C_2) & \A(A_3) & \A(C_3) & \A(A_4) & \A(C_4)\\
 \midrule
(\wf:Z=0) & \{u_1\langle\rangle,u_2\langle\rangle\} & \{u_1\langle \rangle,u_2\langle\rangle\} & \{u_1\langle\rangle,u_2\langle\rangle\} & \{u_1\langle\rangle,u_2\langle\rangle\} & \{u_3\langle\rangle,u_5\langle\rangle\} & \{u_3\langle\rangle,u_5\langle\rangle\} & \{u_3\langle\rangle,u_4\langle\rangle\} & \{u_3\langle\rangle,u_4\langle\rangle\}\\
(\wf:P_1^S=8)  & \{u_1\langle\rangle,u_2\langle\rangle\} & \{u_1\langle \rangle,u_2\langle\rangle\} & \{u_1\langle\rangle,u_2\langle\rangle\} & \{u_1\langle\rangle,u_2\langle\rangle\} & \{u_3\langle\rangle,u_5\langle\rangle\} & \{u_3\langle\rangle,u_5\langle\rangle\} & \{u_3\langle\rangle,u_4\langle\rangle\} & \{u_3\langle\rangle,u_4\langle\rangle\}\\

(u_2:A_1=9) & \{u_1\langle\rangle,u_2\langle\rangle\} & \mathbf{\{u_1\langle C_1, \leq \rangle,u_2\langle\rangle\}} & \{u_1\langle\rangle,u_2\langle\rangle\} & \{u_1\langle\rangle,u_2\langle\rangle\} & \{u_3\langle\rangle,u_5\langle\rangle\} & \{u_3\langle\rangle,u_5\langle\rangle\} & \{u_3\langle\rangle,u_4\langle\rangle\} & \{u_3\langle\rangle,u_4\langle\rangle\}\\
(u_2:C_1=12) & \{u_1\langle\rangle,u_2\langle\rangle\} & \mathbf{\{u_1\langle 12, \leq \rangle,u_2\langle\rangle\}} & \mathbf{\{u_1\langle\rangle,u_2\langle 14, \leq \rangle\}} & \{u_1\langle\rangle,u_2\langle\rangle\} & \{u_3\langle\rangle,u_5\langle\rangle\} & \{u_3\langle\rangle,u_5\langle\rangle\} & \{u_3\langle\rangle,u_4\langle\rangle\} & \{u_3\langle\rangle,u_4\langle\rangle\}\\
(\wf:P_2^S=15) & \{u_1\langle\rangle,u_2\langle\rangle\} & \{u_1\langle 12, \leq \rangle,u_2\langle\rangle\} & \{u_1\langle\rangle,u_2\langle 14, \leq \rangle\} & \{u_1\langle\rangle,u_2\langle\rangle\} & \{u_3\langle\rangle,u_5\langle\rangle\} & \{u_3\langle\rangle,u_5\langle\rangle\} & \{u_3\langle\rangle,u_4\langle\rangle\} & \{u_3\langle\rangle,u_4\langle\rangle\}\\
(\wf:P_3^S=15) & \{u_1\langle\rangle,u_2\langle\rangle\} & \{u_1\langle 12, \leq \rangle,u_2\langle\rangle\} & \{u_1\langle\rangle,u_2\langle 14, \leq \rangle\} & \{u_1\langle\rangle,u_2\langle\rangle\} & \{u_3\langle\rangle,u_5\langle\rangle\} & \{u_3\langle\rangle,u_5\langle\rangle\} & \{u_3\langle\rangle,u_4\langle\rangle\} & \{u_3\langle\rangle,u_4\langle\rangle\}\\
(u_2:A_2=15) & \{u_1\langle\rangle,u_2\langle\rangle\} & \{u_1\langle 12, \leq \rangle,u_2\langle\rangle\} & \{u_1\langle\rangle,u_2\langle 14, \leq \rangle\} & \mathbf{\{u_1\langle C_2, \leq \rangle,u_2\langle\rangle\}} & \{u_3\langle\rangle,u_5\langle\rangle\} & \{u_3\langle\rangle,u_5\langle\rangle\} & \{u_3\langle\rangle,u_4\langle\rangle\} & \{u_3\langle\rangle,u_4\langle\rangle\}\\
(u_2:C_2=20) & \{u_1\langle\rangle,u_2\langle\rangle\} & \{u_1\langle 12, \leq \rangle,u_2\langle\rangle\} & \{u_1\langle\rangle,u_2\langle 14, \leq \rangle\} & \mathbf{\{u_1\langle 20, \leq \rangle,u_2\langle\rangle\}} & \{u_3\langle\rangle,u_5\langle\rangle\} & \{u_3\langle\rangle,u_5\langle\rangle\} & \{u_3\langle\rangle,u_4\langle\rangle\} & \{u_3\langle\rangle,u_4\langle\rangle\}\\
(\wf:P_1^E=20) & \{u_1\langle\rangle,u_2\langle\rangle\} & \{u_1\langle 12, \leq \rangle,u_2\langle\rangle\} & \{u_1\langle\rangle,u_2\langle 14, \leq \rangle\} & \{u_1\langle 20, \leq \rangle,u_2\langle\rangle\} & \{u_3\langle\rangle,u_5\langle\rangle\} & \{u_3\langle\rangle,u_5\langle\rangle\} & \{u_3\langle\rangle,u_4\langle\rangle\} & \{u_3\langle\rangle,u_4\langle\rangle\}\\

(\wf:B^S=21) & \{u_1\langle\rangle,u_2\langle\rangle\} & \{u_1\langle 12, \leq \rangle,u_2\langle\rangle\} & \{u_1\langle\rangle,u_2\langle 14, \leq \rangle\} & \{u_1\langle 20, \leq \rangle,u_2\langle\rangle\} & \{u_3\langle\rangle,u_5\langle\rangle\} & \{u_3\langle\rangle,u_5\langle\rangle\} & \{u_3\langle\rangle,u_4\langle\rangle\} & \{u_3\langle\rangle,u_4\langle\rangle\}\\

(\wf:B^E=21) & \{u_1\langle\rangle,u_2\langle\rangle\} & \{u_1\langle 12, \leq \rangle,u_2\langle\rangle\} & \{u_1\langle\rangle,u_2\langle 14, \leq \rangle\} & \{u_1\langle 20, \leq \rangle,u_2\langle\rangle\} & \{u_3\langle\rangle,u_5\langle\rangle\} & \{u_3\langle\rangle,u_5\langle\rangle\} & \{u_3\langle\rangle,u_4\langle\rangle\} & \{u_3\langle\rangle,u_4\langle\rangle\}\\

(u_3:A_3=22) & \{u_1\langle\rangle,u_2\langle\rangle\} & \{u_1\langle 12, \leq \rangle,u_2\langle\rangle\} & \{u_1\langle\rangle,u_2\langle 14, \leq \rangle\} & \{u_1\langle 20, \leq \rangle,u_2\langle\rangle\} & \{u_3\langle\rangle,u_5\langle\rangle\} & \mathbf{\{u_3\langle\rangle,u_5\langle C_3, \leq \rangle\}} & \mathbf{\{u_3\langle C_3, \leq \rangle,u_4\langle\rangle\}} & \{u_3\langle\rangle,u_4\langle\rangle\}\\
(u_4:A_4=22) & \{u_1\langle\rangle,u_2\langle\rangle\} & \{u_1\langle 12, \leq \rangle,u_2\langle\rangle\} & \{u_1\langle\rangle,u_2\langle 14, \leq \rangle\} & \{u_1\langle 20, \leq \rangle,u_2\langle\rangle\} & \{u_3\langle\rangle,u_5\langle\rangle\} & \{u_3\langle\rangle,u_5\langle C_3, \leq \rangle\} & \{u_3\langle C_3, \leq \rangle,u_4\langle\rangle\} & \mathbf{\{u_3\langle\rangle,u_4\langle C_4, \leq \rangle\}}\\

(u_3:C_3=23) & \{u_1\langle\rangle,u_2\langle\rangle\} & \{u_1\langle 12, \leq \rangle,u_2\langle\rangle\} & \{u_1\langle\rangle,u_2\langle 14, \leq \rangle\} & \{u_1\langle 20, \leq \rangle,u_2\langle\rangle\} & \{u_3\langle\rangle,u_5\langle\rangle\} & \mathbf{\{u_3\langle\rangle,u_5\langle 23, \leq \rangle\}} & \mathbf{\{u_3\langle 23, \leq \rangle,u_4\langle\rangle\}} & \{u_3\langle\rangle,u_4\langle C_4, \leq \rangle\}\\
(\wf:P_2^E=24) & \{u_1\langle\rangle,u_2\langle\rangle\} & \{u_1\langle 12, \leq \rangle,u_2\langle\rangle\} & \{u_1\langle\rangle,u_2\langle 14, \leq \rangle\} & \{u_1\langle 20, \leq \rangle,u_2\langle\rangle\} & \{u_3\langle\rangle,u_5\langle\rangle\} & \{u_3\langle\rangle,u_5\langle 23, \leq \rangle\} & \{u_3\langle 23, \leq \rangle,u_4\langle\rangle\} & \{u_3\langle\rangle,u_4\langle C_4, \leq \rangle\}\\
(u_4:C_4=25) & \{u_1\langle\rangle,u_2\langle\rangle\} & \{u_1\langle 12, \leq \rangle,u_2\langle\rangle\} & \{u_1\langle\rangle,u_2\langle 14, \leq \rangle\} & \{u_1\langle 20, \leq \rangle,u_2\langle\rangle\} & \{u_3\langle\rangle,u_5\langle\rangle\} & \{u_3\langle\rangle,u_5\langle 23, \leq \rangle\} & \{u_3\langle 23, \leq \rangle,u_4\langle\rangle\} & \mathbf{\{u_3\langle\rangle,u_4\langle 25, \leq \rangle\}}\\

(\wf:E^S=26) & \{u_1\langle\rangle,u_2\langle\rangle\} & \{u_1\langle 12, \leq \rangle,u_2\langle\rangle\} & \{u_1\langle\rangle,u_2\langle 14, \leq \rangle\} & \{u_1\langle 20, \leq \rangle,u_2\langle\rangle\} & \{u_3\langle\rangle,u_5\langle\rangle\} & \{u_3\langle\rangle,u_5\langle 23, \leq \rangle\} & \{u_3\langle 23, \leq \rangle,u_4\langle\rangle\} & \{u_3\langle\rangle,u_4\langle 25, \leq \rangle\}\\

(\wf:E^E=26) & \{u_1\langle\rangle,u_2\langle\rangle\} & \{u_1\langle 12, \leq \rangle,u_2\langle\rangle\} & \{u_1\langle\rangle,u_2\langle 14, \leq \rangle\} & \{u_1\langle 20, \leq \rangle,u_2\langle\rangle\} & \{u_3\langle\rangle,u_5\langle\rangle\} & \{u_3\langle\rangle,u_5\langle 23, \leq \rangle\} & \{u_3\langle 23, \leq \rangle,u_4\langle\rangle\} & \{u_3\langle\rangle,u_4\langle 25, \leq \rangle\}\\

(\wf:P_3^E=27) & \{u_1\langle\rangle,u_2\langle\rangle\} & \{u_1\langle 12, \leq \rangle,u_2\langle\rangle\} & \{u_1\langle\rangle,u_2\langle 14, \leq \rangle\} & \{u_1\langle 20, \leq \rangle,u_2\langle\rangle\} & \{u_3\langle\rangle,u_5\langle\rangle\} & \{u_3\langle\rangle,u_5\langle 23, \leq \rangle\} & \{u_3\langle 23, \leq \rangle,u_4\langle\rangle\} & \{u_3\langle\rangle,u_4\langle 25, \leq \rangle\}\\
\bottomrule
\end{tabular}}

\scalebox{0.55}{\begin{tabular}{*{12}{>{$}l<{$}}}
\toprule

\texttt{Executed TP (cont.)} & \A(Z) & \A(P_1^S) & \A(P_1^E) & \A(P_2^S) & \A(P_2^E) & \A(P_3^S) & \A(P_3^E) & \A(B^S) & \A(B^E) & \A(E^S) & \A(E^E) \\
\midrule
(\wf:Z=0) &  \{\wf\langle\rangle\} & \{\wf\langle\rangle\} & \{\wf\langle\rangle\} & \{\wf\langle\rangle\} & \{\wf\langle\rangle\} & \{\wf\langle\rangle\} & \{\wf\langle\rangle\} & \{\wf\langle\rangle\} & \{\wf\langle\rangle\} & \{\wf\langle\rangle\} & \{\wf\langle\rangle\} \\
(\wf:P_1^S=8) & \{\wf\langle\rangle\} & \{\wf\langle\rangle\} & \{\wf\langle\rangle\} & \{\wf\langle\rangle\} & \{\wf\langle\rangle\} & \{\wf\langle\rangle\} & \{\wf\langle\rangle\} & \{\wf\langle\rangle\} & \{\wf\langle\rangle\} & \{\wf\langle\rangle\} & \{\wf\langle\rangle\} \\
(u_2:A_1=9) & \{\wf\langle\rangle\} & \{\wf\langle\rangle\} & \{\wf\langle\rangle\} & \{\wf\langle\rangle\} & \{\wf\langle\rangle\} & \{\wf\langle\rangle\} & \{\wf\langle\rangle\} & \{\wf\langle\rangle\} & \{\wf\langle\rangle\} & \{\wf\langle\rangle\} & \{\wf\langle\rangle\} \\
(u_2:C_1=12) & \{\wf\langle\rangle\} & \{\wf\langle\rangle\} & \{\wf\langle\rangle\} & \{\wf\langle\rangle\} & \{\wf\langle\rangle\} & \{\wf\langle\rangle\} & \{\wf\langle\rangle\} & \{\wf\langle\rangle\} & \{\wf\langle\rangle\} & \{\wf\langle\rangle\} & \{\wf\langle\rangle\} \\
(\wf:P_2^S=15) & \{\wf\langle\rangle\} & \{\wf\langle\rangle\} & \{\wf\langle\rangle\} & \{\wf\langle\rangle\} & \{\wf\langle\rangle\} & \{\wf\langle\rangle\} & \{\wf\langle\rangle\} & \{\wf\langle\rangle\} & \{\wf\langle\rangle\} & \{\wf\langle\rangle\} & \{\wf\langle\rangle\} \\
(\wf:P_3^S=15) & \{\wf\langle\rangle\} & \{\wf\langle\rangle\} & \{\wf\langle\rangle\} & \{\wf\langle\rangle\} & \{\wf\langle\rangle\} & \{\wf\langle\rangle\} & \{\wf\langle\rangle\} & \{\wf\langle\rangle\} & \{\wf\langle\rangle\} & \{\wf\langle\rangle\} & \{\wf\langle\rangle\} \\
(u_2:A_2=15) & \{\wf\langle\rangle\} & \{\wf\langle\rangle\} & \{\wf\langle\rangle\} & \{\wf\langle\rangle\} & \{\wf\langle\rangle\} & \{\wf\langle\rangle\} & \{\wf\langle\rangle\} & \{\wf\langle\rangle\} & \{\wf\langle\rangle\} & \{\wf\langle\rangle\} & \{\wf\langle\rangle\} \\
(u_2:C_2=20) & \{\wf\langle\rangle\} & \{\wf\langle\rangle\} & \{\wf\langle\rangle\} & \{\wf\langle\rangle\} & \{\wf\langle\rangle\} & \{\wf\langle\rangle\} & \{\wf\langle\rangle\} & \{\wf\langle\rangle\} & \{\wf\langle\rangle\} & \{\wf\langle\rangle\} & \{\wf\langle\rangle\} \\
(\wf:P_1^E=20) & \{\wf\langle\rangle\} & \{\wf\langle\rangle\} & \{\wf\langle\rangle\} & \{\wf\langle\rangle\} & \{\wf\langle\rangle\} & \{\wf\langle\rangle\} & \{\wf\langle\rangle\} & \{\wf\langle\rangle\} & \{\wf\langle\rangle\} & \{\wf\langle\rangle\} & \{\wf\langle\rangle\} \\
(\wf:B^S=21) & \{\wf\langle\rangle\} & \{\wf\langle\rangle\} & \{\wf\langle\rangle\} & \{\wf\langle\rangle\} & \{\wf\langle\rangle\} & \{\wf\langle\rangle\} & \{\wf\langle\rangle\} & \{\wf\langle\rangle\} & \{\wf\langle\rangle\} & \{\wf\langle\rangle\} & \{\wf\langle\rangle\} \\
(\wf:B^E=21) & \{\wf\langle\rangle\} & \{\wf\langle\rangle\} & \{\wf\langle\rangle\} & \{\wf\langle\rangle\} & \{\wf\langle\rangle\} & \{\wf\langle\rangle\} & \{\wf\langle\rangle\} & \{\wf\langle\rangle\} & \{\wf\langle\rangle\} & \{\wf\langle\rangle\} & \{\wf\langle\rangle\} \\
(u_3:A_3=22) & \{\wf\langle\rangle\} & \{\wf\langle\rangle\} & \{\wf\langle\rangle\} & \{\wf\langle\rangle\} & \{\wf\langle\rangle\} & \{\wf\langle\rangle\} & \{\wf\langle\rangle\} & \{\wf\langle\rangle\} & \{\wf\langle\rangle\} & \{\wf\langle\rangle\} & \{\wf\langle\rangle\} \\
(u_4:A_4=22) & \{\wf\langle\rangle\} & \{\wf\langle\rangle\} & \{\wf\langle\rangle\} & \{\wf\langle\rangle\} & \{\wf\langle\rangle\} & \{\wf\langle\rangle\} & \{\wf\langle\rangle\} & \{\wf\langle\rangle\} & \{\wf\langle\rangle\} & \{\wf\langle\rangle\} & \{\wf\langle\rangle\} \\
(u_3:C_3=23) & \{\wf\langle\rangle\} & \{\wf\langle\rangle\} & \{\wf\langle\rangle\} & \{\wf\langle\rangle\} & \{\wf\langle\rangle\} & \{\wf\langle\rangle\} & \{\wf\langle\rangle\} & \{\wf\langle\rangle\} & \{\wf\langle\rangle\} & \{\wf\langle\rangle\} & \{\wf\langle\rangle\} \\
(\wf:P_2^E=24) & \{\wf\langle\rangle\} & \{\wf\langle\rangle\} & \{\wf\langle\rangle\} & \{\wf\langle\rangle\} & \{\wf\langle\rangle\} & \{\wf\langle\rangle\} & \{\wf\langle\rangle\} & \{\wf\langle\rangle\} & \{\wf\langle\rangle\} & \{\wf\langle\rangle\} & \{\wf\langle\rangle\} \\
(u_4:C_4=25) & \{\wf\langle\rangle\} & \{\wf\langle\rangle\} & \{\wf\langle\rangle\} & \{\wf\langle\rangle\} & \{\wf\langle\rangle\} & \{\wf\langle\rangle\} & \{\wf\langle\rangle\} & \{\wf\langle\rangle\} & \{\wf\langle\rangle\} & \{\wf\langle\rangle\} & \{\wf\langle\rangle\} \\
(\wf:E^S=26) & \{\wf\langle\rangle\} & \{\wf\langle\rangle\} & \{\wf\langle\rangle\} & \{\wf\langle\rangle\} & \{\wf\langle\rangle\} & \{\wf\langle\rangle\} & \{\wf\langle\rangle\} & \{\wf\langle\rangle\} & \{\wf\langle\rangle\} & \{\wf\langle\rangle\} & \{\wf\langle\rangle\} \\
(\wf:E^E=26) & \{\wf\langle\rangle\} & \{\wf\langle\rangle\} & \{\wf\langle\rangle\} & \{\wf\langle\rangle\} & \{\wf\langle\rangle\} & \{\wf\langle\rangle\} & \{\wf\langle\rangle\} & \{\wf\langle\rangle\} & \{\wf\langle\rangle\} & \{\wf\langle\rangle\} & \{\wf\langle\rangle\} \\
(\wf:P_3^E=27) & \{\wf\langle\rangle\} & \{\wf\langle\rangle\} & \{\wf\langle\rangle\} & \{\wf\langle\rangle\} & \{\wf\langle\rangle\} & \{\wf\langle\rangle\} & \{\wf\langle\rangle\} & \{\wf\langle\rangle\} & \{\wf\langle\rangle\} & \{\wf\langle\rangle\} & \{\wf\langle\rangle\} \\

\bottomrule
\end{tabular}}
\vspace*{0.2cm}\end{table*}

\begin{EXT}
\section{Proofs}

\begin{proof}[Lemma~\ref{lem:actask}]

Consider an STNU fragment consisting of one contingent link $A
\xRightarrow[]{[x,y]} C$ connected to a requirement link $P_{n+1,z}^S
\xrightarrow[]{[k,k]} P_{n+1,z}^E$ by means of the mapping given in
Table~\ref{tbl:connRule} for some chosen $P$, $n \in \mathbb{N}$ and $z \in
\mathit{Displacement}(P)$. Also, we know that $k = \mathit{Granularity}(P)$
and $k < y$ (by assumption). The STNU fragment along with its labeled distance
graph are depicted in Figure~\ref{fig:lemma1:STNU} and
Figure~\ref{fig:lemma1:LDG}, respectively.

The controllability algorithm \cite{DBLP:conf/aaai/MorrisM05} iteratively
checks for the consistency of the labeled distance graph $\mathit{AllMax}$
projection. This projection must remain consistent during the whole execution
which in every cycle adds new edges to the labeled distance graph according
to its edge generations rules to make explicit the restriction to the
execution strategies \cite{DBLP:conf/aaai/MorrisM05}. If the
$\mathit{AllMax}$ projection survives (i.e., if it remains consistent once
the algorithm has terminated), then the STNU is dynamically controllable.

To get the $\mathit{AllMax}$ projection (Figure~\ref{fig:lemma1:AllMax}), we
simply remove from the labeled distance graph all the \emph{lower-case} edges
and all the upper-case labels from the \emph{upper-case} ones. Of course, this
projection represents the situation in which the contingent links take their
maximum duration (in this case, $\omega = (y)$). Moreover, since by definition
the range $[x,y]$ of each contingent link $A \Rightarrow C$ respects $0 < x <
y < \infty$, it follows that $-y < -x$. Thus, since in the $\mathit{AllMax}$
projection we have both $C \xrightarrow[]{-x} A$ and $C \xrightarrow[]{-y} A$,
and we know that $-y$ is tighter than $-x$, we just keep $C \xrightarrow[]{-y}
A$.


Now, we know by assumption that $k < y$. It turns out that the cycle
$P_{n+1,z}^S \xrightarrow[]{k} P_{n+1,z}^E \xrightarrow[]{0} C
\xrightarrow[]{-y} A \xrightarrow[]{0} P_{n+1,z}^S $ has negative weight that
is initially detected by the consistency checking (carried out by Johnson's
algorithm~\cite{Cormen:2009:IAT:1614191}) that computing the all pairs
shortest paths algorithm answers the question of whether or not the
$\mathit{AllMax}$ projection is consistent. If it is not (as in this case),
the algorithm has failed. Consequently, the initial STNU is not dynamically
controllable because the inconsistency of $\mathit{AllMax}$ projection
excludes \emph{weak controllability}, which in turn, excludes \emph{dynamic
controllability}~\cite{DBLP:journals/jetai/VidalF99}.
\end{proof}

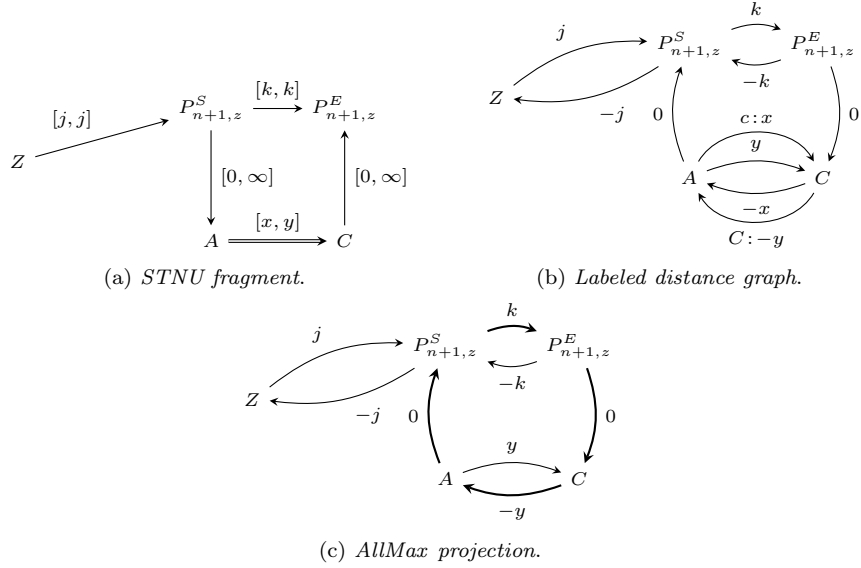
\begin{figure*}[t]\centering
\subfloat[][\emph{STNU fragment}.\label{fig:lemma1:STNU}]{
\begin{tikzpicture}[node distance=50pt,auto]
  \node (Z) {$Z$};
  \node (PS) [right=of Z,,yshift=20pt]{$P_{n+1,z}^S$};
  \node (PE) [right of=PS] {$P_{n+1,z}^E$};
  \node (A) [below of=PS] {$A$};
  \node (C) [below of=PE] {$C$};
  
  \draw[Link]  (Z) to node {} node [] {$[j,j]$} (PS);    
  \draw[Link]  (PS) to node {} node [] {$[k,k]$} (PE);    
  \draw[Link]  (PS) to node {} node [] {$[0,\infty]$} (A);    
  \draw[Link]  (C) to node {} node [swap] {$[0,\infty]$} (PE);    
  \draw[ContingentLink]  (A) to node {} node [] {$[x,y]$} (C);    
\end{tikzpicture}}\qquad
\subfloat[][\emph{Labeled distance graph}.\label{fig:lemma1:LDG}]{
\begin{tikzpicture}[node distance=50pt,auto]
  \node (Z) {$Z$};
  \node (PSlabeled) [right=of Z,yshift=20pt] {$P_{n+1,z}^S$};
  \node (PElabeled) [right of=PSlabeled] {$P_{n+1,z}^E$};
  \node (Alabeled) [below of=PSlabeled] {$A$};
  \node (Clabeled) [below of=PElabeled] {$C$};

  \draw[Link,bend left=20]  (Z) to node {} node [] {$j$} (PSlabeled);
  \draw[Link,bend left=20]  (PSlabeled) to node {} node [] {$-j$} (Z);
  
  \draw[Link,bend left=20]  (PSlabeled) to node {} node [] {$k$} (PElabeled);
  \draw[Link,bend left=20]  (PElabeled) to node {} node [] {$-k$} (PSlabeled);
  \draw[Link,bend left=20]  (Alabeled) to node {} node [] {$0$} (PSlabeled);
  \draw[Link,bend left=20]  (PElabeled) to node {} node [] {$0$} (Clabeled);
  \draw[Link,bend left=20]  (Alabeled) to node {} node [] {$y$} (Clabeled);
  \draw[Link,bend left=20]  (Clabeled) to node {} node [] {$-x$} (Alabeled);
  \draw[Link,bend left=60]  (Alabeled) to node {} node [] {$c\!:\!x$} (Clabeled);
  \draw[Link,bend left=60]  (Clabeled) to node {} node [] {$C\!:\!-y$} (Alabeled);
\end{tikzpicture}}\qquad
\subfloat[][\emph{$\mathit{AllMax}$ projection}.\label{fig:lemma1:AllMax}]{\begin{tikzpicture}[node distance=50pt,auto]
  \node (Z) {$Z$};
  \node (PSlabeled) [right=of Z,yshift=20pt] {$P_{n+1,z}^S$};
  \node (PElabeled) [right of=PSlabeled] {$P_{n+1,z}^E$};
  \node (Alabeled) [below of=PSlabeled] {$A$};
  \node (Clabeled) [below of=PElabeled] {$C$};
  
  \draw[Link,bend left=20]  (Z) to node {} node [] {$j$} (PSlabeled);
  \draw[Link,bend left=20]  (PSlabeled) to node {} node [] {$-j$} (Z);
  \draw[Link,thick,bend left=20]  (PSlabeled) to node {} node [] {$k$} (PElabeled);
  \draw[Link,bend left=20]  (PElabeled) to node {} node [] {$-k$} (PSlabeled);
  \draw[Link,thick,bend left=20]  (Alabeled) to node {} node [] {$0$} (PSlabeled);
  \draw[Link,thick,bend left=20]  (PElabeled) to node {} node [] {$0$} (Clabeled);
  \draw[Link,bend left=20]  (Alabeled) to node {} node [] {$y$} (Clabeled);
  \draw[Link,thick,bend left=20]  (Clabeled) to node {} node [] {$-y$} (Alabeled);
\end{tikzpicture}}
\caption{Supporting figures for the proof of Lemma~\ref{lem:actask}.}

\end{figure*}

\begin{proof}[Theorem~\ref{thm:pt2stn}]
The mapping $\ptToSTN{} : P \times I \rightarrow \langle \mathcal{T},\allowbreak \mathcal{C} \rangle$ returns an STN starting from a periodic expression $P$ whose applicability is limited by an interval $I$ whose upper bound is $\neq \infty$. Thus, for each generated interval $I_{n+1,z}^P = [t_1,\dots,t_g]$ of integers (where $g = \mathit{Granularity}(P)$) the equivalent interval of real values is $[t_1-1,t_g]$ which is modeled in the STN as an equivalent requirement link $P_{n+1,z}^S \xrightarrow[]{[k,k]} P_{n+1,z}^E$ where $k = t_g - (t_1 -1)$ ($k \geq 0)$. In addition, all of these requirement links are connected to the source node $Z$ (\emph{zero-time point}) $Z \xrightarrow[]{[j,j]} P_{n+1,z}^S$ where $j = t_1 - 1$ ($j \geq 0)$.
Thus, for all $n,z$ such that $I_{n+1,z}^P \subseteq I$, this mapping builds the STN in Figure~\ref{fig:thm1:STN}.

We are now ready to prove that the resulting STN is (i) consistent, and (ii)
admits exactly one solution.

(i) To prove that the resulting STN is consistent we consider its distance graph depicted in Figure~\ref{fig:thm1:DG}.

Let $w : E \to \mathbb{R}$ be the weight function that maps each edge $(u,v)
\in E$ to a real value in $\mathbb{R}$. Thus, for each pair of
\emph{connected} nodes $u,v$ we know that $w(u,v) = -w(v,u)$ (as each range
of each time point has the form $[x,x]$). Thus, the weight of the cycle $u
\to v \to u$ is $w(u,v) - w(v,u) = 0$. It is quite simple to see that if each
cycle between two connected nodes has weight $0$, so does any other cycle
involving $n > 2$ nodes. It turns out, that there does not exist any cycle
whose weight is negative, thus the STN is consistent because the all pairs
shortest paths algorithm terminates correctly.
 
(ii) The reason why the STN admits one and only one solution is due to the
fact that for each pair of connected nodes $u,v$ we know that $w(u,v) =
-w(v,u)$. Consequently, the weight of each path from a source node $s$ to any
other node $u$ is equal to the negation of that going from $u$ to $s$. From
\cite{DBLP:conf/kr/DechterMP89} we know that the weight of the shortest path
from $s$ to $u$ is the upper bound of the range of allowed values for the
time point $u$, whereas the negation of the opposite direction (i.e., of the
shortest path from $u$ to $s$) corresponds to the lower bound. When these two
values are equal (note that we negate the negation of the path from $u$ to
$s$), then the range of the allowed values for time point $u$ collapses to
only one point. When this holds for all time points, the STN has exactly one
solution since the range of allowed values of each time point consists of one
possible value only.
\end{proof}
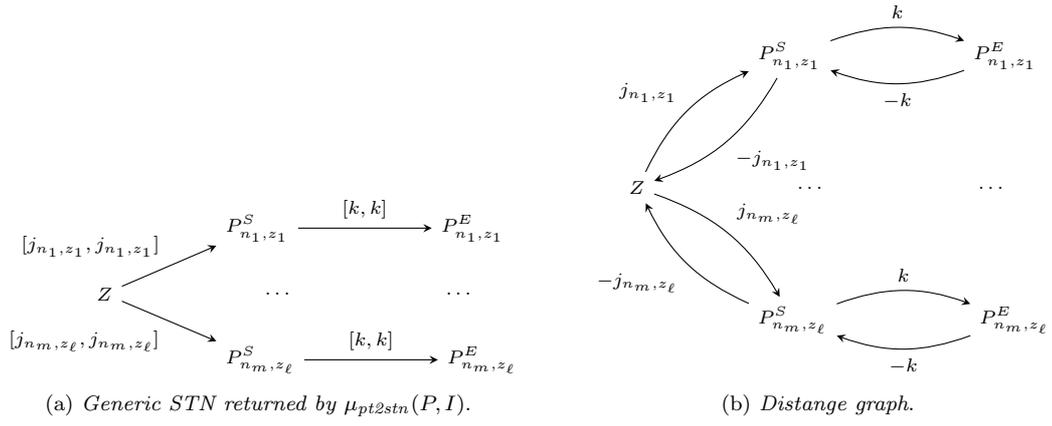
\begin{figure*}[t]\centering
\subfloat[][\emph{Generic STN returned by $\ptToSTN{}(P,I)$}.\label{fig:thm1:STN}]{
\begin{tikzpicture}[node distance=50pt]
  \node (Z) {$Z$};
  \node (P1n1z1S) [above right=of Z,yshift=-25pt] {$P_{n_1,z_1}^S$};
  \node (P1n1z1E) [right=of P1n1z1S] {$P_{n_1,z_1}^E$};
  \node (PnzS) [right=of Z] {\dots};
  \node (PnzE) [right=of PnzS] {\dots};

  \node (P1nMzS) [below right=of Z,yshift=25pt] {$P_{n_{m},z_\ell}^S$};
  \node (P1nMzE) [right=of P1nMzS] {$P_{n_{m},z_\ell}^E$};

  \draw[Link]  (Z) -- node [] {$[j_{n_1,z_1},j_{n_1,z_1}]$} (P1n1z1S);
  \draw[Link]  (P1n1z1S) to node {} node [] {$[k,k]$} (P1n1z1E);
  \draw[Link]  (Z) -- node [swap] {$[j_{n_{m},z_\ell},j_{n_{m},z_\ell}]$} (P1nMzS);   
  \draw[Link]  (P1nMzS) to node {} node [] {$[k,k]$} (P1nMzE);    
\end{tikzpicture}}\qquad
\subfloat[][\emph{Distange graph}.\label{fig:thm1:DG}]{
\begin{tikzpicture}[node distance=50pt]
  \node (Z) {$Z$};
  \node (P1n1z1S) [above right=of Z] {$P_{n_1,z_1}^S$};
  \node (P1n1z1E) [right=of P1n1z1S] {$P_{n_1,z_1}^E$};
  \node (PnzS) [right=of Z] {\dots};
  \node (PnzE) [right=of PnzS] {\dots};
  \node (P1nMzS) [below right=of Z] {$P_{n_{m},z_\ell}^S$};
  \node (P1nMzE) [right=of P1nMzS] {$P_{n_{m},z_\ell}^E$};

  \draw [->,bend left=20] (Z) to node {$j_{n_1,z_1}$} node [] {} (P1n1z1S);
  \draw [->,bend left=20] (P1n1z1S) to node {$-j_{n_1,z_1}$} node [] {} (Z);

  \draw[Link,bend left=20]  (P1n1z1S) to node {} node [] {$k$} (P1n1z1E);
  \draw[Link,bend left=20]  (P1n1z1E) to node {} node [] {$-k$} (P1n1z1S);
  

  \draw [->,bend left=20] (Z) to node {$j_{n_{m},z_\ell}$} node [] {} (P1nMzS);
  \draw [->,bend left=20] (P1nMzS) to node {$-j_{n_{m},z_\ell}$} node [] {} (Z);

  \draw[Link,bend left=20]  (P1nMzS) to node {} node [] {$k$} (P1nMzE);   
  \draw[Link,bend left=20]  (P1nMzE) to node {} node [] {$-k$} (P1nMzS);    
\end{tikzpicture}}
\caption{Supporting figures for the proof of Theorem~\ref{thm:pt2stn}.}
\end{figure*}

\end{EXT}


\end{document}